\documentclass[journal=jpccck,manuscript=article]{achemso}
\SectionNumbersOn

\usepackage[version=3]{mhchem} 
\usepackage[T1]{fontenc}       
\usepackage{float} 
\usepackage{esvect} 
\usepackage{color,soul} 
\usepackage{subcaption} 
\usepackage{color} 

\author{Shima Alizadeh}
\affiliation{Department of Mechanical Engineering, 
Flow Physics and Computational Engineering,
Stanford University, 
Stanford, California 94305, USA}
\altaffiliation{Center for Turbulence Research, Stanford University, Stanford, California 94305, USA}

\author{Martin Z. Bazant}
\affiliation{Department of Chemical Engineering and Department of Mathematics, 
Massachusetts Institute of Technology, 
Cambridge, Massachusetts 02139, USA} 

\author{Ali Mani}
\alsoaffiliation{Department of Mechanical Engineering, 
Flow Physics and Computational Engineering,
Stanford University, 
Stanford, California 94305, USA}
\altaffiliation{Center for Turbulence Research, Stanford University, Stanford, California 94305, USA}
\email{alimani@stanford.edu}
\phone{+1650 7251325}
\fax{+1650 7253525}

\title
  {Impact of Network Heterogeneity on Nonlinear Electrokinetic Transport in Porous Media}


\begin{document}


\begin{abstract}
We present a numerical study of nonlinear electrokinetic transport in porous media, focusing on the role of heterogeneity in a porous microstructure on ion concentration polarization and over-limiting current. For simplicity, the porous medium is modeled as a network of long, thin charged cylindrical pores, each governed by one-dimensional effective transport equations. For weak surface conduction, when sufficiently large potential is applied, we demonstrate that electrokinetic transport in a porous network can be dominated by electroconvection via internally induced flow loops, which is not properly captured by existing homogenized models. We systematically vary the topology and ``accessivity'' of the pore network and compare with simulations of traditional homogenized parallel-pore (capillary-bundle) models, in order to reveal the effects of regular and hierarchical connectivity. Our computational framework sheds light on the complex physics of nonlinear electrokinetic phenomena in microstructures and may be used to design porous media for applications, such as water desalination and purification by shock electrodialysis. 
\end{abstract}

\section{Introduction}
Nonlinear electrokinetic phenomena, such as ion concentration polarization (ICP) \cite{probstein2003,nikonenko2010}, electro-osmotic instability and electro-convection ~\cite{zaltzman2007,rubinstein2000,rubinstein2008,yossifon2008,kim2007,kwak2013,zaltzman2017,nikonenko2017effect}, over-limiting current (OLC)~\cite{nikonenko2014,dydek2011,yaroshchuk_coupled_2011,rubinstein2013,nielsen2014, nam2015}, deionization shock waves~\cite{manimartin2011,mani2009,zangle2009,zangle2010,yaroshchuk2012acis}, current-induced membrane discharge~\cite{anderson2012}, and induced-charge electro-osmosis~\cite{cocis2010, martin2009_acis},  have mostly been studied in simple geometries, defined by planar membranes or electrodes, isolated colloidal particles, or straight microfluidic channels.  On the other hand, emerging applications, such as capacitive deionization~\cite{porada2013} and ion separation~\cite{zhao2012}, electro-osmotic pumping \cite{suss2011}, resistive switching~\cite{han2016resistive}, electrokinetic fingering~\cite{mirzadeh2017}, shock electrodialysis~\cite{schlumpberger2015,deng2013,deng2015}, and shock electrodeposition~\cite{han2014,han2016dendrite}, naturally involve heterogeneous porous media, which leverage nanoscale electrokinetic phenomena at engineering scales more easily and effectively than is possible with microfluidic devices\cite{sheridan2011, knust2013, kim2010, crooks2008, crooks2009}. As such, there is a need to develop mathematical theories and simulation methods that go beyond classical ``leaky membrane'' ~\cite{dydek2013,manimartin2011,yaroshchuk2012acis,spiegler1971} and ``porous electrode''~\cite{newman_book,biesheuvel2010,biesheuvel2011} homogenized models to capture the complex physics of nonlinear electrokinetics in porous microstructures. In the absence of fluid flow, surface conduction (SC) has recently been shown to depend on the microstructure of porous electrodes~\cite{mirzadeh2014}, but to our knowledge, the coupled effects of SC and electrokinetic flows in complex porous networks have not yet been analyzed, despite their being implicated in experimental observations of OLC and deionization shock waves in porous media~\cite{deng2013,han2014,han2016dendrite}.   Here, we leverage our recently developed computational framework for nonlinear electrokinetics in porous media~\cite{alizadeh1, alizadeh2} to begin answering these fundamental questions related to OLC in charged porous media.

Diffusion-limited current is a general property of electrochemical systems~\cite{newman_book}, which separates distinct operating regimes in processes ranging from electrodialysis~\cite{nikonenko2014} to electrodeposition~\cite{rosso2007, bai2017}.  For steady conduction through a dilute, symmetric, binary bulk electrolyte from a reservoir to an ideally cation-selective membrane or electrode surface, the limiting current scales as $I_{\text{lim}} = \frac{2zeDC_{\text{ref}}}{L}A$, in which $C_{\text{ref}}$ is the reservoir concentration and $z, e$  and $D$ are ion valence, elementary charge, and ion diffusivity respectively. As discussed in the supplementary information, the dimensional diffusion-limited current ($\tilde{I}_{\text{dl}}$) in this prototypical case of ICP exhibits a diode-like dependence on the applied voltage ($V$) across electrolyte~\cite{probstein2003,levich1947,dydek2011}:
\begin{equation}
 \tilde{I}_{\text{dl}} = I_{\text{lim}} (1 - \exp(-V/V_T)),
\label{eq:diffusion_limited_current}
\end{equation}
where $V_T=\frac{k_B T}{\text{z}e}$ is the thermal voltage, $k_B$ is Boltzmann constant, and $T$ is absolute temperature. (Note that throughout this paper we distinguish dimensional variables from dimensionless values by using ``\~{ }'' on top of the symbols associated with dimensional variables.) 
The diffusion-limited current can only be exceeded (leading to OLC) if the preceding assumptions break down, by providing additional ions (via water splitting~\cite{nikonenko2014,kniaginicheva2015water} or membrane discharge~\cite{anderson2012}), activating new transport mechanisms (such as electro-osmotic instability~\cite{zaltzman2017,nikonenko2014}, surface conduction or electro-osmotic flow~\cite{dydek2011,yaroshchuk_coupled_2011,deng2013,nam2015}), or effectively shortening the diffusion length (by dendritic growth instability~\cite{bai2017,han2014,han2016dendrite}).   

If electrolyte transport is confined laterally by charged surfaces in a microchannel or porous medium during the passage of OLC, then nonlinear dynamics such as deionization shocks \cite{manimartin2011, mani2009, zangle2009, zangle2010} may occur, which can be interpreted as shock waves in the concentration field. Deionization shocks can propagate in porous medium and trigger higher electric current \cite{manimartin2011,deng2013, dydek2011,  dydek2013, yaroshchuk2012acis} than $I_{\text{lim}}$. 
As first predicted by Dydek et al.~\cite{dydek2011}, the current-voltage relation is nearly linear above the diffusion-limited regime across a range of experiments and theoretical models \cite{deng2013,nam2015,dydek2013,yaroshchuk2012acis,nielsen2014,han2014,han2016dendrite,han2016resistive}:
\begin{equation}
I = I_{\text{dl}} + \sigma_{\text{OLC}} V/V_T,
\end{equation}
where $\sigma_{\text{OLC}}$ is normalized overlimiting conductance (scaled to $I_{\text{lim}}$), but the scaling of  $\sigma_{\text{OLC}}$ with properties of the electrolyte and confining geometry has not yet been fully understood. 

There have been a number of studies that aimed to understand the physical mechanisms that sustain OLC in electrokinetic systems. The basic mechanism of OLC sustained by electro-osmotic convection in a microchannels was first proposed by Yaroshchuk et al.~\cite{yaroshchuk_coupled_2011}, albeit using a Taylor-Aris  model for hydrodynamic dispersion in the pressure-driven backflow that balances electro-osmotic forward flow, which only applies to very long  microchannels.   Dydek at al. ~\cite{dydek2011} first predicted theoretically that OLC in a micro-channel can be dominated by one of three fundamental mechanisms -- SC through electric double layers (EDLs), electro-osmosis (EO) along the sidewalls or electro-osmotic instability at the end surface, depending on the geometry and surface properties -- and they proposed a modified scaling theory of OLC by EO convection. The predicted transition between the SC and EO mechanisms was later confirmed by Nam et al. ~\cite{nam2015} in microfluidic experiments, consistent with the predicted scalings of $\sigma_{\text{OLC}}$ with channel height in each regime. A  few years earlier, Deng et al. \cite{deng2013} conducted the first experiments to investigate the contributing mechanisms to OLC in porous media. Their measurements revealed an approximate a power law dependence of $\sigma_{\text{OLC}}$ on reservoir concentration ($C_{\text{ref}}$) and pH-regulated surface charge $q_s$, which was consistent with theoretical prediction of Dydek et al. \cite{dydek2011}: $\sigma_{\text{OLC}}\sim q_s^{2/5} c_0^{4/5}$. This agreement was remarkable given that the theory of Dydek et al. was not derived for a porous material, but rather a single 2D micro-channel having only internal flow recirculation near the wall region. The general effect of surface convection leading to mixing during OLC in a microchannel was later studied by Rubinstein and Zaltzman~\cite{rubinstein2013}, but again without considering any porous network effects.   In order to rationalize the experimental results, Deng et al. \cite{deng2013} hypothesized that the scaling of Dydek must also hold at the network scale, based on eddies recirculating around loops in the pore network, rather than within individual pores. 

While OLC and its mechanisms have been studied comprehensively in a micro-channel setting \cite{dydek2011,yaroshchuk_transport_2011,yaroshchuk_coupled_2011,rubinstein2013,nielsen2014,nam2015} to the best of our knowledge there has not been any theoretical study of nonlinear electrokinetic phenomena in porous structures, going beyond macroscopic ``leaky membrane'' continuum models~\cite{manimartin2011,yaroshchuk2012acis,dydek2013}. More specifically, the impact of network heterogeneity and the induced flow on OLC and its connection with SC in porous structures have never been investigated through a rigorous numerical setting. In this article, we present numerical simulations of electrokinetic transport in porous structures that involve random networks of micro-scale and nano-scale pores, which appear to be the first computational investigations of nonlinear electrokinetic phenomena in porous networks. 

We employ our previously developed computational framework \cite{alizadeh1, alizadeh2} to conduct the simulations of electrokinetic transport in massive random networks of pores. In particular, we examine the internally induced flows due to random pore coupling and connectivity loops and investigate how such flows can lead to enhancement of OLC in systems subject to ICP. As a measure of OLC, we report the I-V characteristics computed for a variety of porous networks and compare them to those obtained from corresponding homogenized models. As described in Section (\ref{sec:lattice_model_problem}), homogenized models consist of decoupled straight pores and have the same total volume to surface area ratio (the same surface conduction) as the random porous networks. To develop comprehensive quantification we explore the impacts of network topology and pore size randomness, as well as electrolyte salinity on the overlimiting conductance. Lastly, we assess the applicability of the previously proposed theories \cite{deng2013,dydek2011} based on the electroosmotic flow mechanism for OLC in porous materials.

\section{Problem Setup}
\subsection{Computational Framework} \label{sec:computational framework}
In our previous two-paper series \cite{alizadeh1, alizadeh2}, we presented comprehensive derivation and validation of our multi-scale model that enables pore-scale simulation of nonlinear electrokinetic phenomena in porous structures. We model a porous structure as a network of micro-scale and nano-scale pores, each is to be long and thin. We solve non-dimensionalized area-integrated transport equation for coin species:
\begin{equation}
S(x)\frac{\partial \overline{C^-}}{\partial t} +\frac{\partial}{\partial x} \{ S(x) \overline{F^-} \}=0,
\label{eq:area_neg_ion}
\end{equation} 
where $x$ coordinate is in pore's axial direction, $S(x)$ is the local pore area cross-section, and $ \overline{C^-}$ and $\overline{F^-}$ represent respectively the area-averaged concentration of coions and the area-averaged flux of coins. In our formulation, we write $\overline{F^-}$ in terms of local gradients of three driving potentials multiplied with pre-factors that capture the effects of local EDL and surface charge density:
\begin{equation*}
 \overline{F^-}= \frac{\kappa}{2 \lambda_D^2} \{ \overline{C^-} \overline{g^p} + C_s \overline{g^{p-}} \} \frac{d P_0}{d x} + \{ \kappa ( \overline{C^-} \overline{g^e} + C_s  \overline{g^{e-}} )+ \overline{C^-}\} \frac{d \mu^+}{d x} 
\end{equation*}
\begin{equation}
+ \{ \frac{\kappa}{2 \lambda_D^2} (\overline{C^-} \overline{g^c} + C_s \overline{g^{c-}}) - 2\bar{g}\}\frac{d C_0}{d x}.
\label{eq:flux}
\end{equation}
We use the cross-sectional invariants of the three driving potentials, which are virtual total pressure, $P_0$, virtual counterion electro-chemical potential, $\mu^+$, and virtual ion concentration $C_0$. Detailed explanation of this choice is presented in our two-paper series \cite{alizadeh1, alizadeh2}. $\bar{g}$, $\overline{g^e}$, $\overline{g^c}$, $\overline{g^{p-}}$, $\overline{g^{e-}}$, $\overline{g^{c-}}$ are the area-averaged coefficients that are tabulated with respect to the dimensionless EDL thickness, $\lambda^*$, and dimensionless surface charge density $q_s^*$. To close equation (\ref{eq:area_neg_ion}), we compute the fields of $P_0$ and $\mu^+$ by solving coupled conservation equations of fluid mass and electric charge. These equations involve additional area-averaged pre-factors which along with the aforementioned coefficients are tabulated with respect to $\lambda^*$ and $q_s^*$. Thus, when solving the discretized version of equation (\ref{eq:area_neg_ion}), we apply an interpolation procedure to extract the area-averaged coefficients from tabulated quantities based on the local values of $\lambda^*$ and $q_s^*$.

Equation (\ref{eq:flux}) involves a number of dimensionless parameters. $\kappa = \frac{\varepsilon}{\mu D}V_T^2$ is electrohydrodynamic coupling constant. $\lambda_D(x) =  \frac{1}{h_p} \sqrt{\frac{\varepsilon k_B T}{2 \tilde{C}_{\text{ref}} \text{z}^2 e^2}}$ is the ratio of reference Debye length to local effective pore size that is defined as the ratio of pore volume to pore surface area. $C_s= 2 |q_s^*| \lambda_D^2$ represents the excess couterions needed to neutralize surface charge at each local cross-section. $C_s$ indicates the significance of surface conduction effect.

\subsection{Generation of Random Networks of Pores}
\begin{figure}[H]
        \centering
        \begin{subfigure}[H]{0.5\textwidth}
                \includegraphics[width=\textwidth]{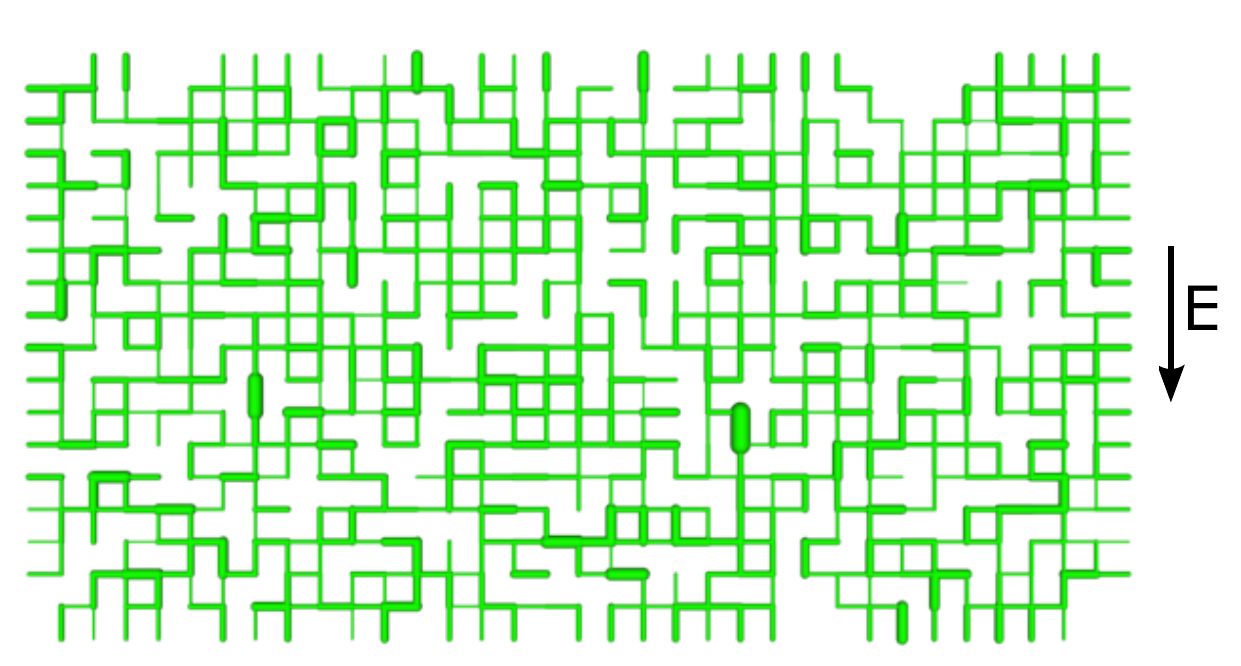}
                \caption{}
                \label{fig:tall_nonhierarch}
        \end{subfigure}%
          
        \begin{subfigure}[H]{0.5\textwidth}
                \includegraphics[width=\textwidth]{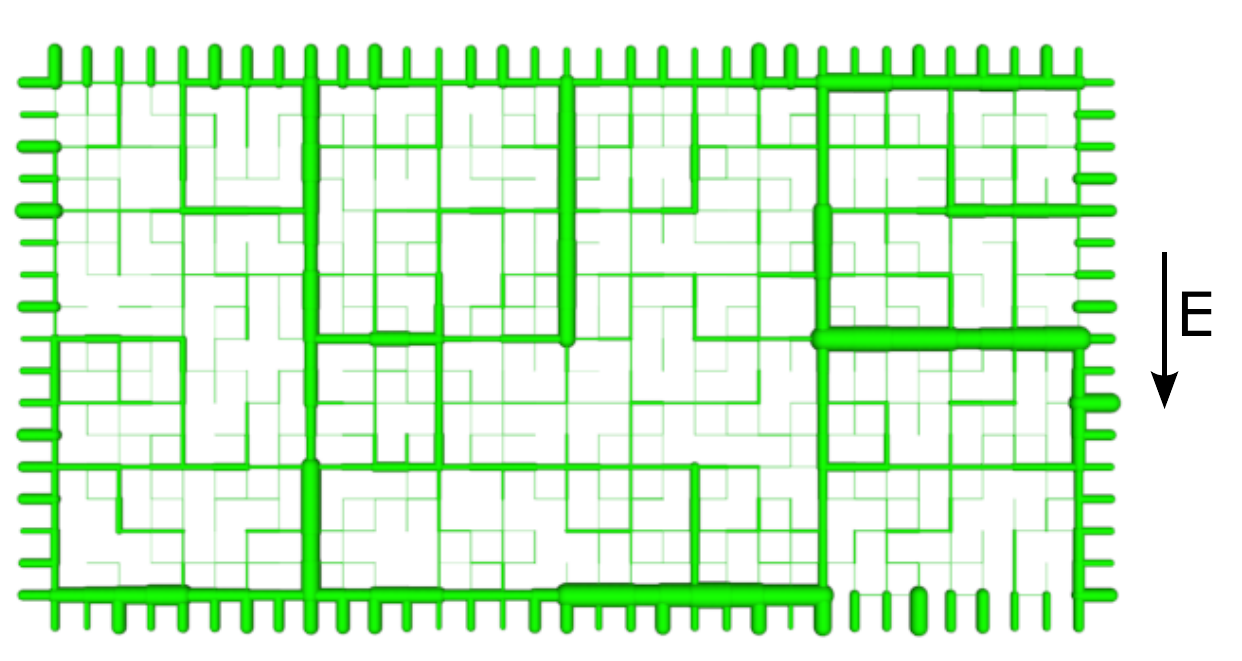}
                \caption{}
                \label{fig:tall_hierarch}
        \end{subfigure}
        ~ 
         \caption{Examples of a non-hierarchical network (a) and a hierarchical network composed of four layers (b). The direction of the external potential gradient is depicted.}
 \label{fig:tall_lattice_media}
  \end{figure}
Although our model does not have any constraint on the topology of a porous medium, we here focus our study on the networks that are constructed on a square background lattice, where pores are coupled horizontally or vertically to each other. We assume all pores have circular cross sections and all have the same length and surface charge density. To enforce the randomness of the network we consider three manipulations: first, we sample pores' diameters from a log-normal distribution for a given mean and standard deviation. Second, we block pores randomly using a uniform random distribution. Pores with probability values below a certain quantity, $p$, are eliminated from the network. Third, we consider hierarchical structures by connecting series of lattices, in which each lattice has a scale twice finer than the previous lattice. In this strategy each lattice is constructed by the two previously mentioned manipulations. The final network is constructed by selecting the pore diameters based on the superposition of all scales. 

In the present study, we consider two types of pore networks with varying degrees of parallel versus series connectivity. The first type has random pore sizes and random pore blocking with only a single-scale lattice, as shown in Figure (\ref{fig:tall_nonhierarch}), and the second is constructed by all three manipulations mentioned above, as shown in Figure (\ref{fig:tall_hierarch}). For the rest of this article, we call the former, ``non-hierarchical network'' and denote the latter by ``hierarchical network''.   For each network topology, we investigate how induced conduction due to network random coupling could contribute to OLC and the propagation of deionization shocks in overlimiting regime. 

The two types of networks correspond to different ranges of ``accessivity'', $\alpha$, a recently proposed parameter that quantifies the degree of parallel versus series connectivity in a finite porous network~\cite{gu2018microscopic}. The classical picture of a ``capillary bundle'' of parallel straight pores corresponds to the limit $\alpha=1$, since every pore is directly connected to the surface, like resistors in parallel.  As we shall demonstrate, our hierarchical networks have intermediate accessivity, which is associated with greatly enhanced electrokinetic convection. In these networks internal pores are connected to the membrane by a chain of one or two pores in series, but there are also loops within the medium, which can support electrokinetic eddy formation.  In contrast, our non-hierarchical regular networks have low accessivity, $\alpha \to 0$, since the number of accessible pores connected to the membrane surface scales with the (small) surface to volume ratio, measured in units of pore length, and pathways through the system involve many changes in pore diameter. The internal pores are thus mostly connected in like resistors in series, although again flow loops are possible, which   
are the focus of this study.  

Although accessivity can be defined and measured experimentally by fitting capillary pressure versus saturation curves during immiscible fluid displacement (as in cyclic mercury porosimetry) to an invasion percolation theory, the measured accessivity can often be approximated by the fraction of pores that are directly connected to the external surface~\cite{gu2018microscopic}, or more generally to a well connected network of larger pores that transmit forces and flows to the representative volume of interest. The latter concept, measuring the fraction of micropores exposed to the external environment via well-connected macropores, has also been used recently to fit hysteretic moisture sorption isotherms (moisture content versus humidity during dry/wetting cycles) in a variety of hierarchical porous media, such as cement paste, porous glasses, carbon black, and dental enamel~\cite{pinson2018inferring,pinson2015hysteresis,jennings2013water}. 

Motivated by these studies, we approximate the accessivity as follows for our different types of networks. For non-hierarchical porous networks, we define the effective accessivity as the ratio of the number of pores connected to the membrane surface (source of the largest electrokinetic forcing entering the network) to the total number of pores. This approximation procedure results in accessivities near zero for non-hierarchical lattices and equal to one for the homogenized capillary-bundle model. For hierarchical networks, due to the existence of multiple layers of pores at different scales, we utilize a slightly different method to estimate the relevant accessivity.  Since transport in hierarchical networks is dominated  by well-connected larger pores, we need to use a definition that captures network heterogeneity at that scale. To this end, we eliminate micro-scale pores, which are located primarily in the finest scale of the hierarchy. Due to higher hydrodynamic resistance, these pores have lower contribution to ion transport towards the membrane and are neglected, and the accessivity is computed from the network of larger pores using the same definition as we used for non-hierarchical networks. In other words, we estimate the accessivity of a hierarchical network by the accessivity of the macropores only, and below we show that this measure of network connectivity is indeed correlated with enhanced electro-osmotic convection.

\subsection{Model Problem and Homogenized Model} \label{sec:lattice_model_problem}

\begin{figure}[H]
        \centering
        \begin{subfigure}[H]{0.45\textwidth}
                \includegraphics[width=\textwidth]{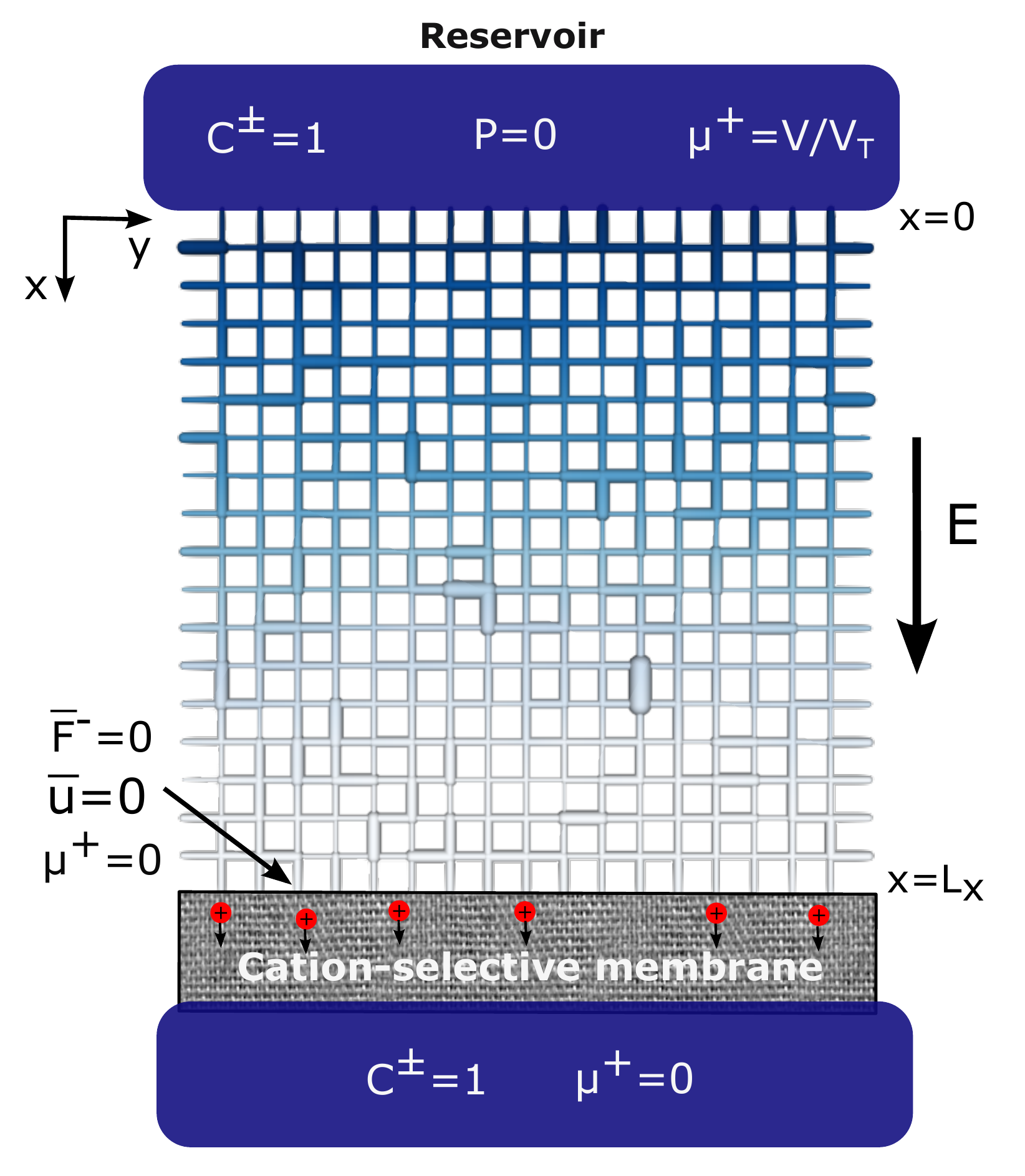}
                \caption{}
                \label{fig:model_lattice}
        \end{subfigure}%
        \begin{subfigure}[H]{0.43\textwidth}
                \includegraphics[width=\textwidth]{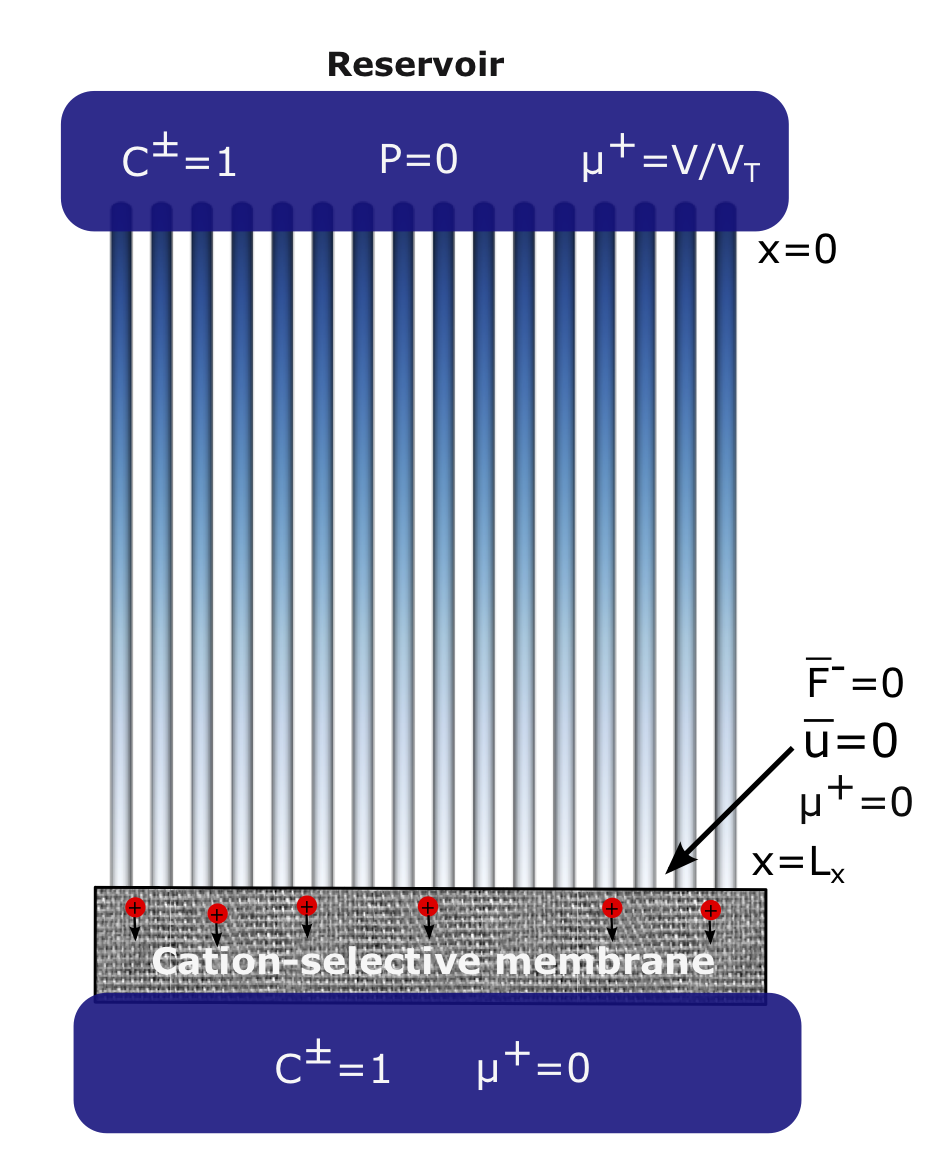}
                \caption{}
                \label{fig:homog_model}
        \end{subfigure}
        ~ 
         \caption{(a) Schematic of a lattice of pores of random diameter confined in between an impermeable cation-exchange-membrane and a well-mixed reservoir. (b) The homogenized capillary-bundle model with the same $h_p$ as the porous network shown in \ref{fig:model_lattice}. Both systems are connected to large reservoirs at x=0 and blocked by a cation-selective membrane at x=$L_x$. There is an external electric field applied downward towards the membrane. The  pore accessivity, approximately equal to the ratio of the number of pores connected to the membrane to the total number of pores, is small ($\alpha=0.06$) for the square lattice fragment shown in (a) and large ($\alpha=1$) for the parallel capillary bundle in (b).} 
 \label{fig:model_problems}
  \end{figure}

Figure (\ref{fig:model_lattice}) shows the schematic of the model problem consisting of a random network of pores confined by an impermeable cation-selective membrane. The membrane imposes zero net flux of anions ($\bar{F}^-(x=L_x)=0$) and zero net flow ($\bar{u}(x=L_x)=0$). The membrane is connected to a big reservoir that contains a symmetric electrolyte with reference concentration of $C_{\text{ref}}$. The electric potential is set to zero in this reservoir. Since the membrane and the reservoir are in equilibrium, the electrochemical potential of cations is nearly constant across the membrane-reservoir interface, as well as across the entire membrane. Thus, the electrochemical potential is fixed and equal to zero at the membrane interface with the random network ($\mu^+(x=L_x)=0$). At x=0 there is a well-mixed reservoir with the same electrolyte concentration as the bottom reservoir. The reservoir pressure is set to atmospheric pressure ($P(x=0)=0$) and the dimensionless applied potential is $V/V_T$, which is equivalent to $\mu^+(x=0)=V/V_T$ at this boundary. Throughout this paper, we report the I-V characteristics with respect to the gradient of electrochemical potential of cations across the random network ( $\Delta\mu^+=V/V_T$).

We report an effective $C_s$ for a random network, which represents the significance of the surface conduction effect in ion transport and is defined based on \textit{network effective pore size}, which can be computed as follows:
\begin{equation}
h_p = \frac{\text{network total volume}}{\text{network total surface}}.
\end{equation}

The effective $C_s$ of a network is then defined as:
\begin{equation}
C_s = -\frac{\sigma}{zeC_{\text{ref}} h_p} = -2q_s^*\lambda_D^2,
\end{equation} 

where, $q_s^*$ and $\lambda_D$ are defined based on the network effective $h_p$. Dydek et al. \cite{dydek2011} demonstrated that for a long and narrow pore the I-V characteristics in units of limiting current and thermal voltage indicates a constant overlimiting conductance equal to pore $C_s$. Later, we report our numerical simulations of electrokinetic transport in random networks of pores and investigate whether there are other mechanisms contributing to overlimiting conductance in these networks.   

The network effective $h_p$ is used to construct an equivalent homogenized (capillary bundle) model of a random network of pores with 100\% accessivity ($\alpha=1$). As shown in Figure (\ref{fig:homog_model}), the homogenized model is composed of parallel replicas of a single straight pore with the same $h_p$ as the network effective $h_p$. Given the results of Dydek et al. \cite {dydek2011} the adapted method of constructing homogenized model ensures that its $C_s$ matches the one for the random network. Therefore, in the absence of mechanisms other than surface conduction we expect that the ratio of overlimiting conductance to limiting current in the homogenized model matches that in the random network. To match the limiting currents, one may manipulate either the number of single pores or their length (or their combination), but since this is a linear operation, it does not provide useful new information. In our computations we simulated a single pore of length, $L_x$, representing the homogenized model but compared I-V responses after normalizing current by the limiting value. Similar to the porous network, the homogenized model is blocked by a cation-exchange membrane at $x=L_x$. Thus, there is no net flow from the reservoir towards the selective membrane in both systems.

We compare the steady state behaviors of the two systems and investigate how pores' random sizes and coupling in a network can influence the system performance against the corresponding homogenized model. 

\subsection{Dimensionless Parameters} \label{sec:dimensionless parameters}
We explore the dependency of overlimiting conductance on the pore blocking probability ($p$), the coefficient of pore size variability $C_v$ defined as the ratio of standard deviation of the pore size distribution to its mean, the effective $C_s$, and the effective $\lambda_D=\frac{\lambda_{\text{ref}}}{h_p}$. The electrolyte electrohydrodynamic coupling constant, $\kappa$, is set to 0.5 to limit the investigation to typical aqueous electrolytes. Table (\ref{tbl:dimless_network}) lists the important dimensionless parameters and the ranges considered in this study.
 
\begin{table}[h]
  \caption{Network dimensionless parameters}
  \label{tbl:dimless_network}
  \centering
  \begin{tabular}{lll}
    \hline
    Dimensionless parameter  & Description  & Range \\
    \hline
    $p$        & Pore blocking probability   & [0, 0.5] \\
    \hline                                                                                                                    
    $C_v$        & ratio of standard deviation to  &  [0, 0.667]\\
                       & mean of the pore size distribution   &    \\
    \hline
    $C_s$       & Network effective surface   conduction & [0.004, 16] \\
    \hline
    $\lambda_D$      &  Network effective Debye length    &  [0.002, 0.5]\\
    \hline
    $V/V_T$  &  Normalized applied potential  &    [0, 60] \\
      \hline
    $\alpha$  &  Accessivity  &    [0.02, 1] \\
  \end{tabular}
\end{table}

The first two parameters, p and $C_v$,  control network geometry, and as a part of this study we explored their influence on system response.  We also examine a wide range of $C_s$ and $\lambda_D$ since the sensitivity to geometry itself may be dependent on these dynamically important parameters. Experimentally, these parameters may be varied by changing the surface material/chemistry and reservoir concentration. As a measure of system response, we consider steady state OLC over a wide range of applied voltages. As we will show, in the overlimiting regime the I-V response can be approximated by a constant slope, and thus we mostly report overlimiting conductance. 

Since for each set of network parameters ($p$, $C_v$, $C_s$, $\lambda_D$, and $\kappa$) we construct the network using random numbers, our numerical results are prone to statistical variations. Only in the limit of very large networks with many pores, this statistical variation becomes negligibly small. In our investigation, we considered larger domain using many pores, as shown in Figure (\ref{fig:tall_lattice_media}). Additionally, for each value of $p$ and $C_v$ we repeat our simulations for a number of random distributions and report the ensemble averaged results for I-V characteristics. In a separate study, to mimic the experimental investigation of Deng et al. \cite{deng2013} we considered variation of dimensional $C_{\text{ref}}$ to investigate the scaling of overlimiting conductance with respect to electrolyte salinity.  

\subsection{Computation of Current} \label{sec:current decomposition}
\begin{figure}[H]
        \centering
        \begin{subfigure}[H]{0.44\textwidth}
                \includegraphics[width=\textwidth]{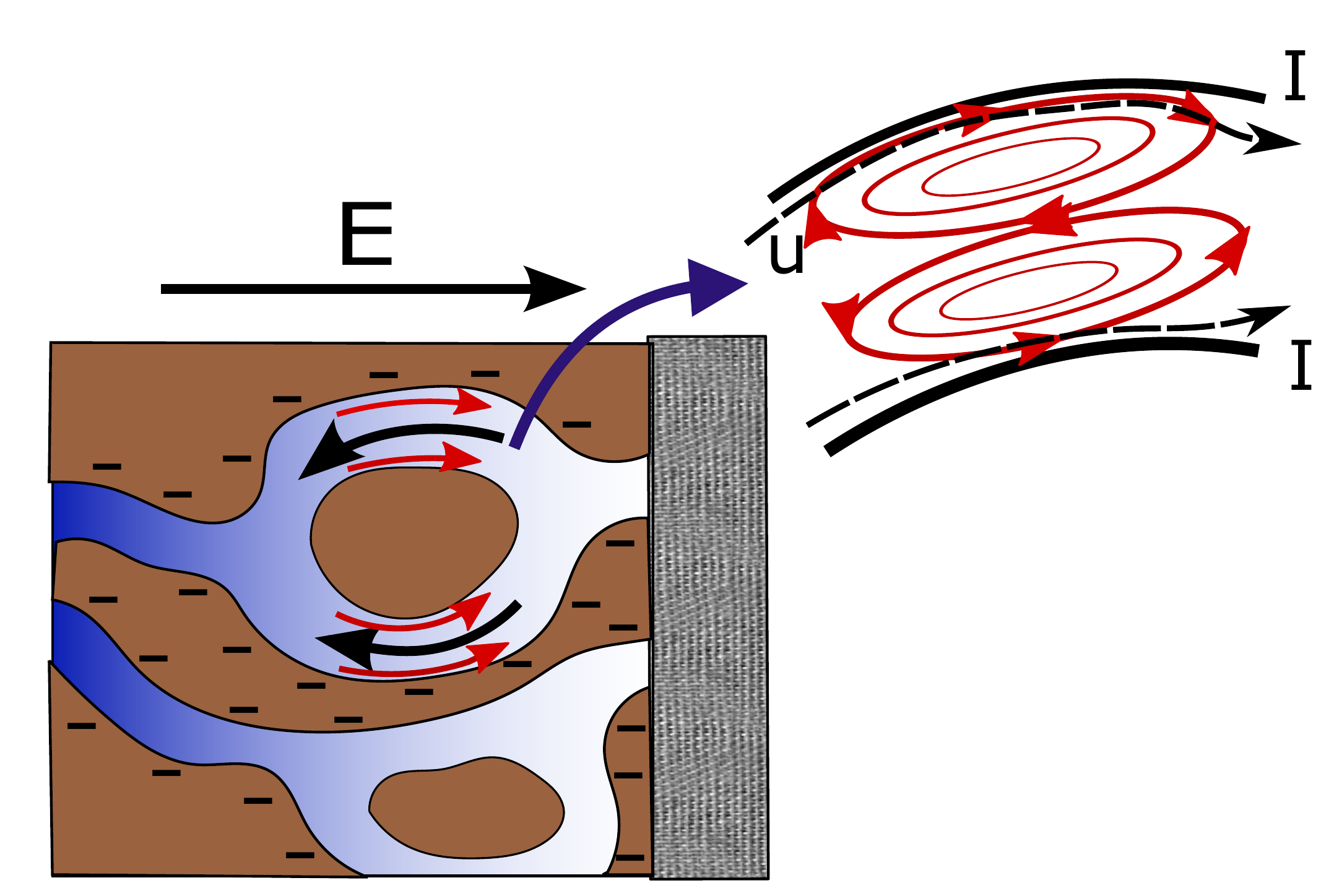}
                \caption{}
                \label{fig:regular}
        \end{subfigure}%
         ~
        \begin{subfigure}[H]{0.44\textwidth}
                \includegraphics[width=\textwidth]{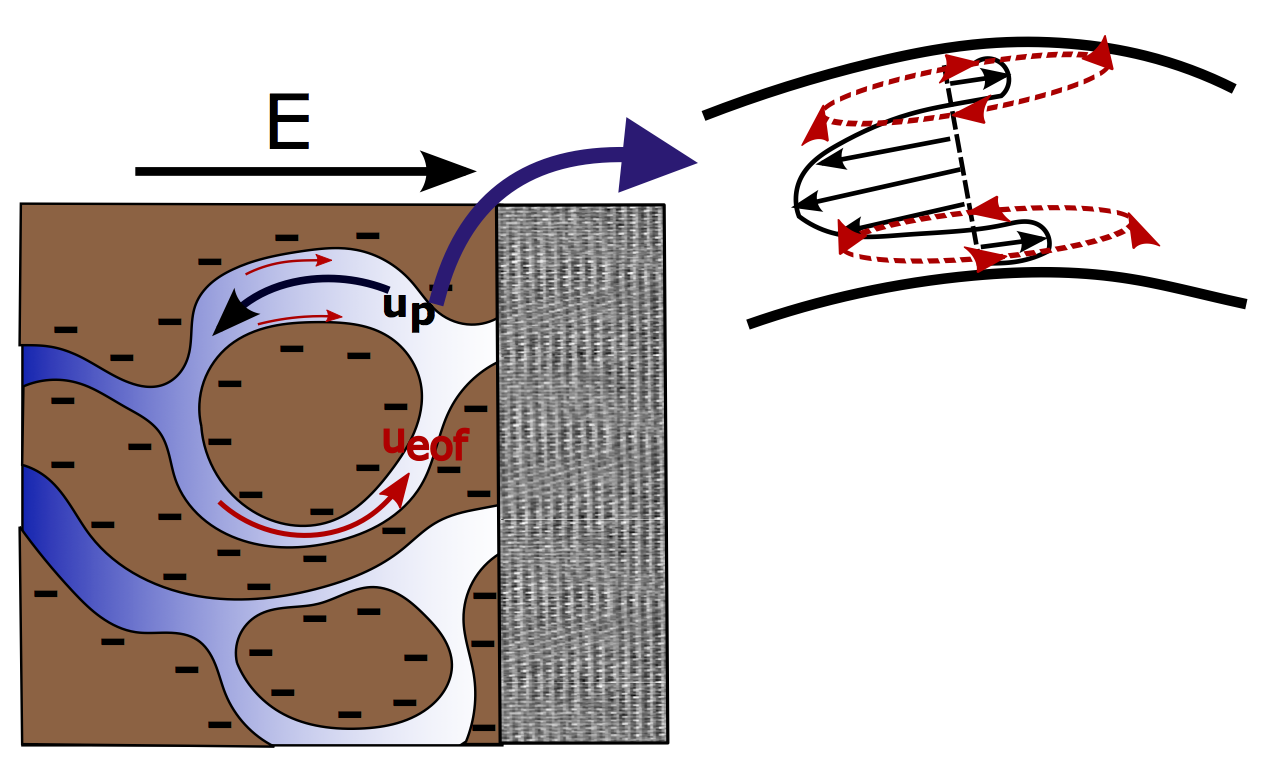}
                \caption{}
                \label{fig:irregular}
        \end{subfigure}
        \caption{(a) Schematic of a local network where the net flow is constrained to be zero within each pore. In such configurations different modes of flow balance each other locally, leading to generation of recirculation zones inside each pore. (b) Schematic of a local network with random pore sizes connecting in a loop. The randomness in the size and the coupling of pores stimulates flow loops across the network in which different modes of flow are globally balanced but with non-zero balance within the pores.}
 \label{fig:two_types_flow}
  \end{figure}
We compute the current of a network at the membrane interface by summing up the current in the pores that are directly connected to the membrane. We report I-V curves normalized by $I_{\text{lim}}$ and $V_T$. $I_{\text{lim}}$ is the maximum amount of charge that can be driven by diffusion mechanism at zero surface charge and zero advection limit. This current can be computed by solving the diffusion equation in the network with zero surface charge with the boundary condition of $C=C_{\text{ref}}$ at the reservoir and $C=0$ at the membrane. While here we use a mathematical trick to compute $I_{\text{lim}}$ by solving diffusion equation with Dirichlet boundary conditions, in practice $I_{\text{lim}}$ can be realized in the limit of high voltages when surface charge is zero and $\lambda_D$ is very small. As shown in Supplementary Information, when the applied voltage is finite, $I_{\text{dl}}$ in equation (\ref{eq:diffusion_limited_current}) can be written in terms of electrochemical potential of counterions as follows:
\begin{equation}
I_{\text{dl}} =  \tilde{I}_{\text{dl}} / I_{\text{lim}} = \left( 1 - \exp(-\Delta \mu^+/2) \right).
\label{eq:diff_limit_i}
\end{equation}

In the settings considered in this study, current above $I_{\text{lim}}$ is expected due to the effects of surface conduction and internally induced flows.  Below, we introduce the idea of current decomposition that we will employ to quantify the impact of surface conduction as well as internally induced flows on the I-V curves. We break down the total current ($I_{\text{tot}}$) into diffusion limited current ($I_{\text{dl}}$) plus components that are driven by surface conduction through EDLs ($I_{\text{sc}}$) and advection effects ($I_{\text{adv}}$). One should note that the advection-driven overlimiting current is composed of two mechanisms, which are shown schematically in Figure (\ref{fig:two_types_flow}). There is an overlimiting current driven by recirculations within pores (Figure(\ref{fig:regular})), which are generated when pressure driven flow and electroosmotic flow balance each other inside a single pore, leading to zero net flow rate. These recirculation zones can exist even in isolated single pores such as those in the homogenized models \cite{deng2013, dydek2011}. The second transport mechanism via advection involves network-level flow loops (Figure(\ref{fig:regular})), which are provoked by the competition between different modes of flow when pores of different sizes are connected in a loop.  

We compute different components of $I_{\text{tot}}$ via the following calculations. First, $I_{\text{tot}}$ is computed by solving the full system of equations without any constraint. Second, the diffusion limitted current, $I_{\text{dl}}$, is computed via equation (\ref{eq:diff_limit_i}) and the computation of $I_{\text{lim}}$ from the solution to the diffusion equation, as discussed above. Thirdly, $I_{\text{dl}}+I_{\text{sc}}$ is computed by turning off all modes of advection. This can be done effectively by setting the tabulated coefficients associated with advection fluxes to zero, and therefore allowing only diffusion and surface conduction modes of transport. $I_{\text{adv}}$ can be computed by subtracting $I_{\text{dl}}+I_{\text{sc}}$ from $I_{\text{tot}}$. Similarly, we denote the  overlimiting conductance due to surface conduction and advection by $\sigma_{\text{olc}, \text{sc}}$ and $\sigma_{\text{olc}, \text{adv}}$ respectively. These conductances can be computed respectively from the slopes of $I_{\text{sc}}$ and $I_{\text{adv}}$ versus the applied voltage. Following Dydek et al. \cite{dydek2011}, we define a conductance ratio $\Gamma = \frac{\sigma_{\text{olc}, \text{adv}}}{\sigma_{\text{olc}, \text{sc}}}$ that indicates the relative importance of advection effect to surface conduction. As we will discuss, for electrokinetic systems with high surface conduction and/or regular structures, $\Gamma < 1$, whereas porous structures with low surface conduction and/or highly random topology can result in $\Gamma\geq1$.

Before proceeding to the result discussion, we would like to point out that our primary objective of this study is to characterize porous materials and examine the impact of network parameters on the character of the I-V curve in the nonlinear regime. Thus, all the I-V curves in our study are normalized by network $I_{\text{lim}}$. We also document the impact of the geometry change on $I_{\text{lim}}$ itself in Supplementary Information. $I_{\text{lim}}$ indicates the ohmic conductance of the material (according to equation (\ref{eq:diff_limit_i})) as $I_{\text{lim}}$ is proportional to $dI/dV$ in the limit of $V=0$. It is an important factor when it comes to the design and the selection of porous materials, and has been examined extensively in the context of percolation and the quantification of the impedance of random geometries subject to Laplace equation and Dirichlet boundary conditions \cite{aharony1985, torquato2002}.Therefore, by documenting $I_{\text{lim}}$ we report solely the impact of geometric manipulation on the ohmic resistance of the network. As noted before, we report ensemble averaged results over 2 or 3 simulations for each set of network parameters. Our results indicate that there is a statistical variation of 1\% to 8\% in $I_{\text{lim}}$. For an accurate quantification more number of ensembles are required to further reduce the statistical error. Nevertheless, in the present analysis the error is sufficiently small to allow capture of the trends intended to be revealed by this study. 
\begin{figure}[H]
        \centering
        \begin{subfigure}[H]{0.45\textwidth}
                \includegraphics[width=\textwidth]{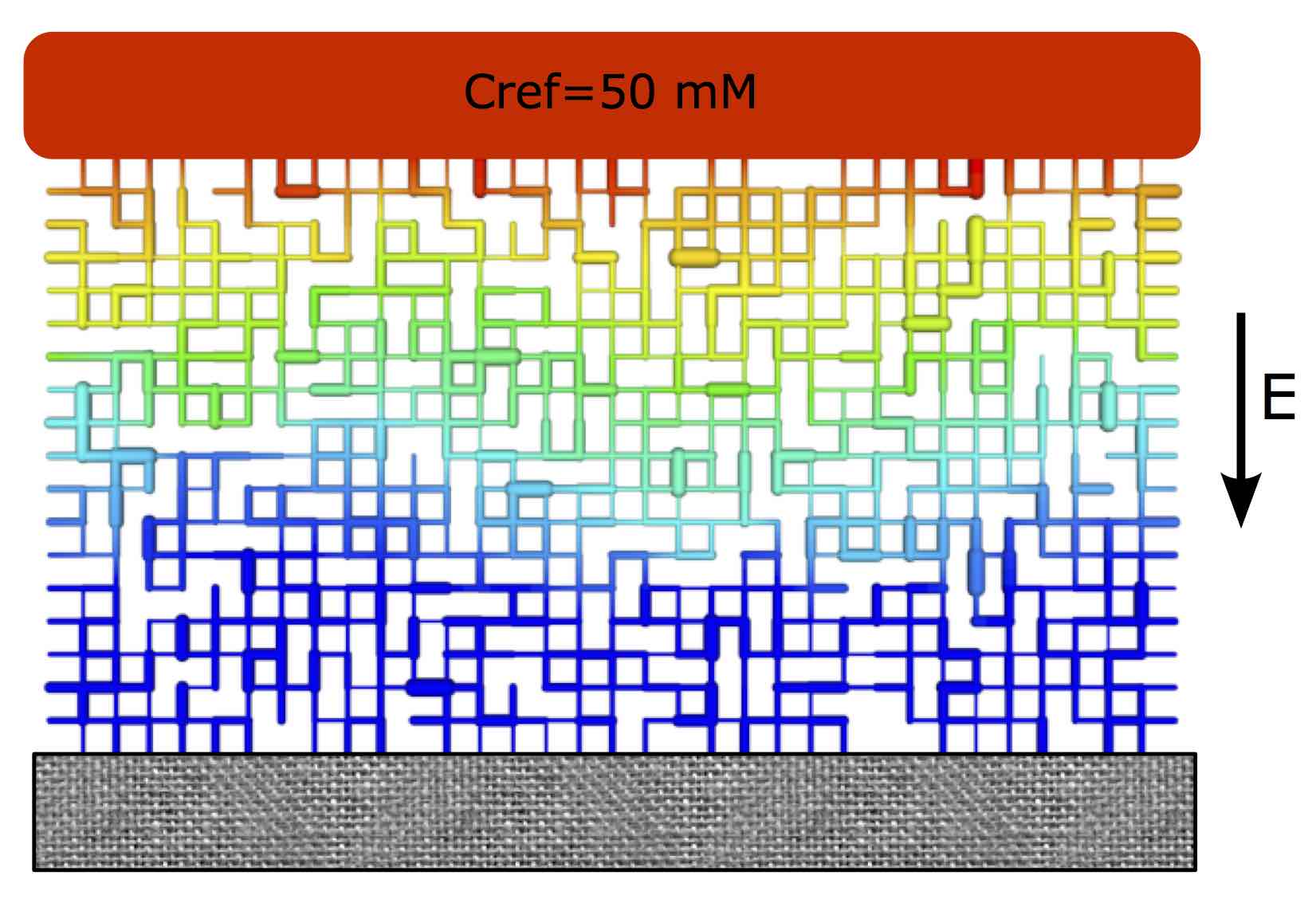}
                \caption{}
                \label{fig:cbar_nonhierarch}
        \end{subfigure}%
        \begin{subfigure}[H]{0.48\textwidth}
                \includegraphics[width=\textwidth]{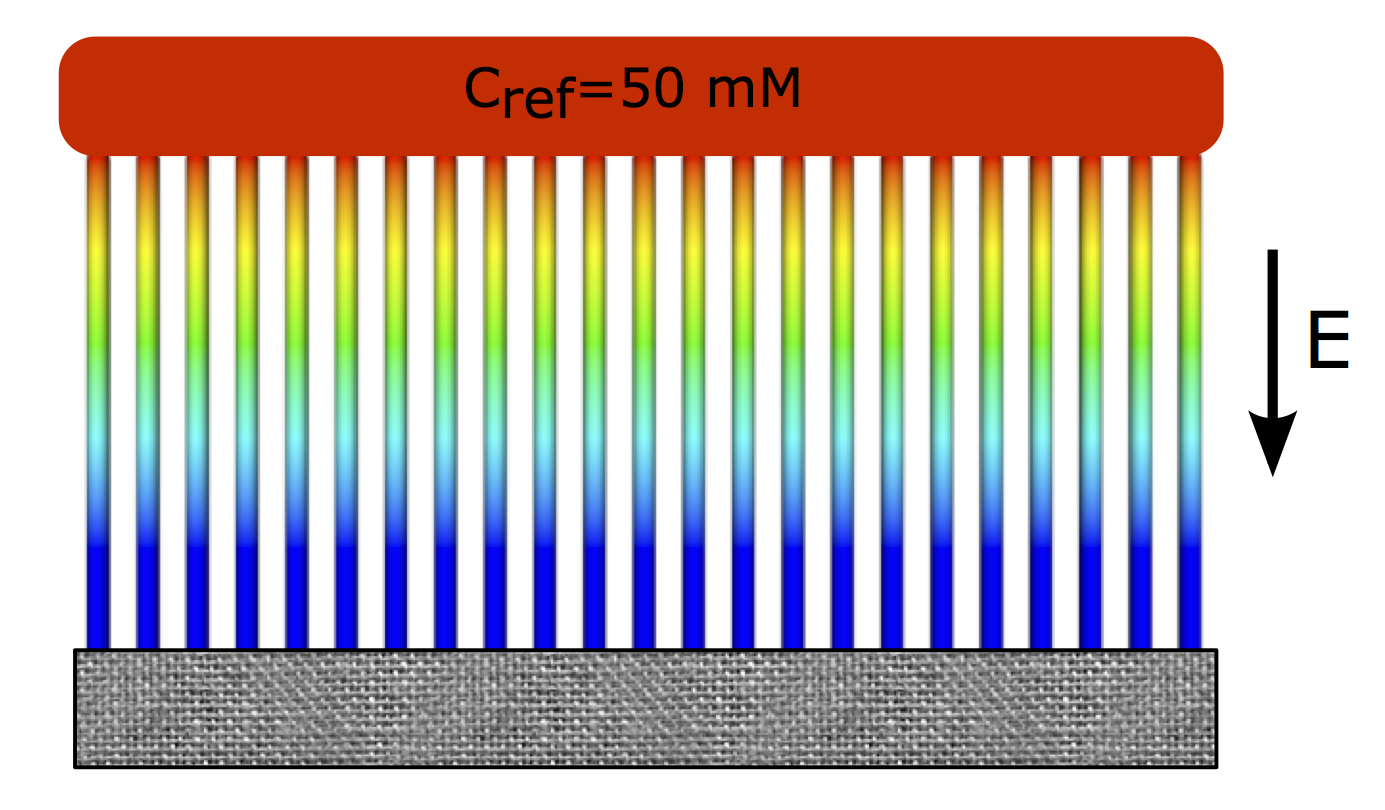}
                \caption{}
                \label{fig:cbar_homog}
        \end{subfigure}
        ~ 
         \caption{(a) The concentration field in a non-hierarchical network exposed to a potential gradient of $V/V_T=50$. (b) The concentration field in the homogenized model which is constructed by replicas of a pore with the same $h_p$ as the effective $h_p$ of the random network.}
\label{fig:nonhierarch_lattice_cfield}
\end{figure}
\section{Results}
\subsection{Electrokinetic Transport in Non-hierarchical Structures} \label{sec:nonhierarchical structures}
Figure (\ref{fig:cbar_nonhierarch}) depicts one example of a non-hierarchical network with low accessivity, $\alpha=0.03$. This network consists  of 936 pores and 633 nodes, and the pore diameters are sampled from a log-normal distribution with mean diameter $d=600 nm$ and standard deviation to mean ratio of $C_v=0.33$. The pore blocking probability is set to $p=0.2$. The random network is connected to a large reservoir that contain KCl solution with $C_{\text{ref}}=50 mM$. We assume that all pores in the network have the same surface charge density equal to $q_s=-10 mC/m^2$. This setting results in the network effective $h_p$ to be $167.06 nm$, which corresponds to the network effective $\lambda_D=0.008$ and $C_s=0.012$. In this regime, each individual pore is thin enough to be in the SC dominated regime~\cite{dydek2011}, so any significant effects of EO convection must involve loops in the porous network.  
\begin{figure}[H]
        \centering
        \begin{subfigure}[H]{0.45\textwidth}
                \includegraphics[width=\textwidth]{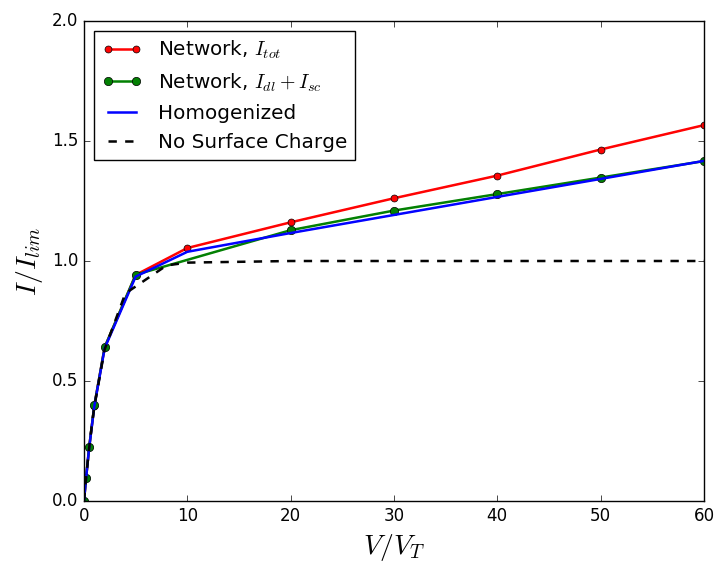}
                \caption{}
                \label{fig:flow_nonhirerachical}
        \end{subfigure}%
        \begin{subfigure}[H]{0.48\textwidth}
                \includegraphics[width=\textwidth]{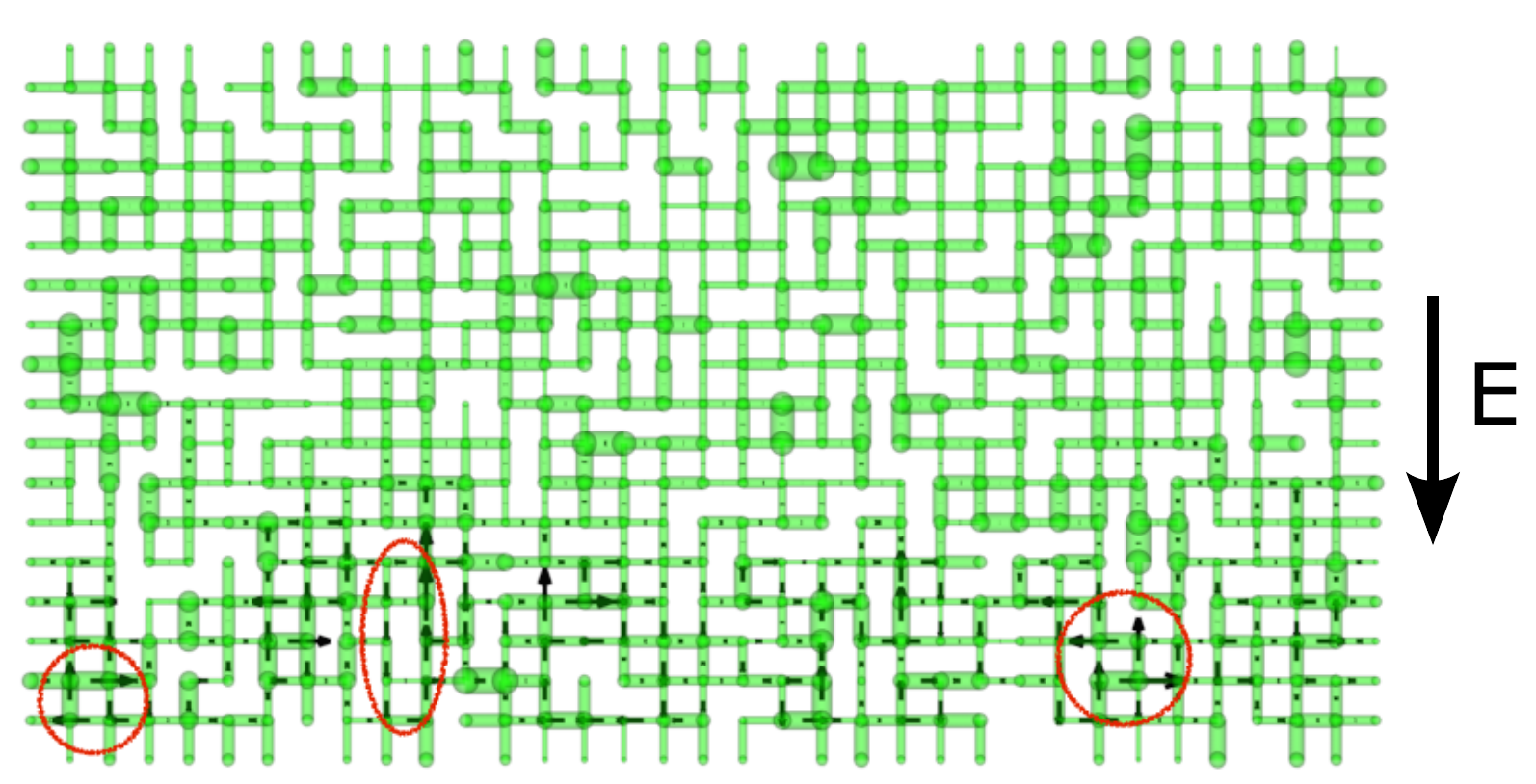}
                \caption{}
                \label{fig:iv_nonhirerachical}
        \end{subfigure}
        ~ 
         \caption{(a) The I-V characteristics of the non-hierarchical random network shown in Figure \ref{fig:cbar_nonhierarch} (red), the same network with flow turned-off (green), the homogenized model (blue), and the diffusion-limited current (dashed black line). (b) The flow field in the non-hierarchical network of Figure \ref{fig:cbar_nonhierarch}. Near the membrane interface, the network involves dominant electroosmotic flow towards the membrane which is balanced by a backward pressure-driven flow, but the effect on OLC is relatively small for this type of porous network.}
\label{fig:iv_nonhirerachical}
\label{fig:nonhierarch_lattice_cfield}
\end{figure}
We used our computational model to study the electrokinetic transport in this network for a wide range of applied voltages. We here present the steady state response of the system for $V/V_T=50$. The concentration fields in the network and the corresponding homogenized model are presented in Figure (\ref{fig:nonhierarch_lattice_cfield}). All pores in the homogenized model have the same $h_p$ equal to 167.06 nm and exhibit SC dominated OLC~\cite{dydek2011,dydek2013,nielsen2014,nam2015}. In the network the random coupling of pores with different pore diameters has caused the deionization shocks to propagate relatively heterogeneously away from the cation-selective membrane, however, the impact of network randomness does not seem to be prominent in this setting. 

Figure (\ref{fig:flow_nonhirerachical}) depicts the induced flow field in the random network. Note that the macroscopic flow field is zero in both random network and the homogenized model. One can see that the internally induced flow exists only near the membrane region, where there is a dominant backward pressure-driven flow, which balances the induced local electroosmotic flows~\cite{yaroshchuk_coupled_2011,dydek2011,deng2013}.   The high electroosmotic flow in the near membrane region results from the higher electric field in this zone. As depicted in Figure (\ref{fig:nonhierarch_lattice_cfield}), this zone has lower concentration due to ICP, and thus requires higher electric field to sustain the same current. The length scale of the flow loops (some loops are marked by red circles on Figure (\ref{fig:flow_nonhirerachical})) are at the scale of pore's length or the length scale of the depletion zone. The induced flow loops have slightly helped the propagation of the deionization shock and made the shock front look nonuniform. Yet, the average impact on conductance in this example is not significant.

Figure(\ref{fig:iv_nonhirerachical}) depicts the I-V curves of the random network and the homogenized model, compared with the diffusion-limited curve or an uncharge material. All I-V curves are normalized by $I_{\text{lim}}$ and $V_T$. For high voltages the random network as well as the homogenized model exhibit overlimiting currents with constant conductances. The random network shows higher overlimiting current with slightly higher conductance, which results from the network random topology and the induced advection effect. The impact is not significant though, as a large portion of the domain involves zero flow and the induced flow loops are only active in a small region close to the membrane. The P{\' e}clet number associated with the flow loops generated in the network for $V/V_T=50$ is $Pe\leq 1$, which demonstrates that the advection effect is not significant in this configuration. To further investigate the contribution of advection to the overlimiting current of the system, we utilized the current decomposition method described in Section (\ref{sec:current decomposition}). The I-V characteristics for this setting (green curve) is close to the one obtained for the homogenized model. Thus, it confirms that the difference between the random network and the homogenized model is primarily due to the induced advection. However, the primary mode of overlimiting transport is surface conduction. We examined a wide variety of non-hierarchical networks using different combinations of $C_v$ and $p$ and the results are presented in the Supplementary Information. Our results demonstrate that despite a finite advection effect induced in all the non-hierarchical networks, surface conduction is the dominant mechanism of OLC with $\Gamma<1$ for all non-hierarchical networks of thin (locally SC dominated) charged pores.

\subsection{Electrokinetic Transport in Hierarchical Structures}  \label{sec:sample_hierarach_results}
\begin{figure}[H]
        \centering
        \begin{subfigure}[H]{0.4\textwidth}
                \includegraphics[width=\textwidth]{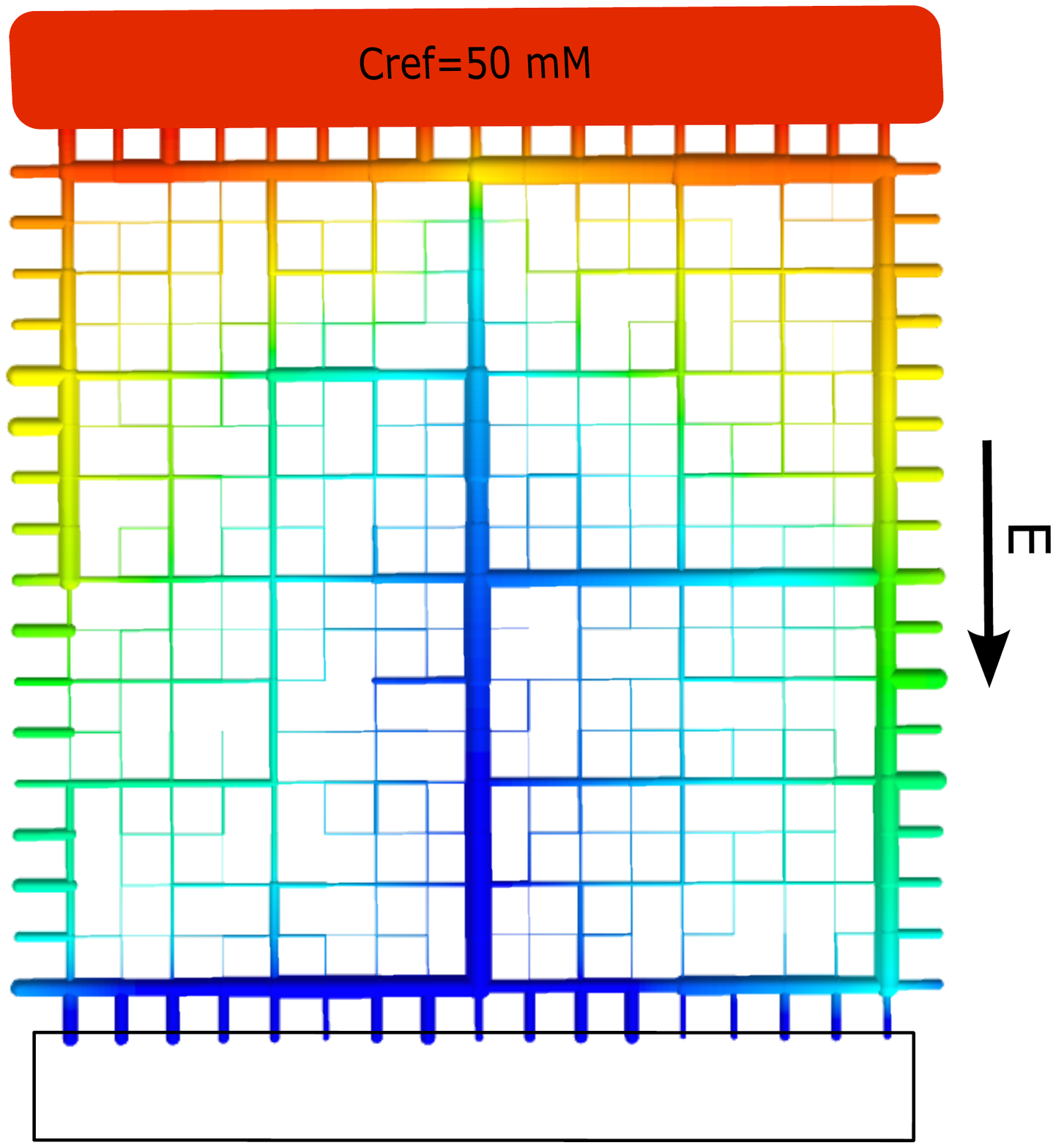}
                \caption{}
                \label{fig:hetero_schematic}
        \end{subfigure}%
          ~
        \begin{subfigure}[H]{0.42\textwidth}
                \includegraphics[width=\textwidth]{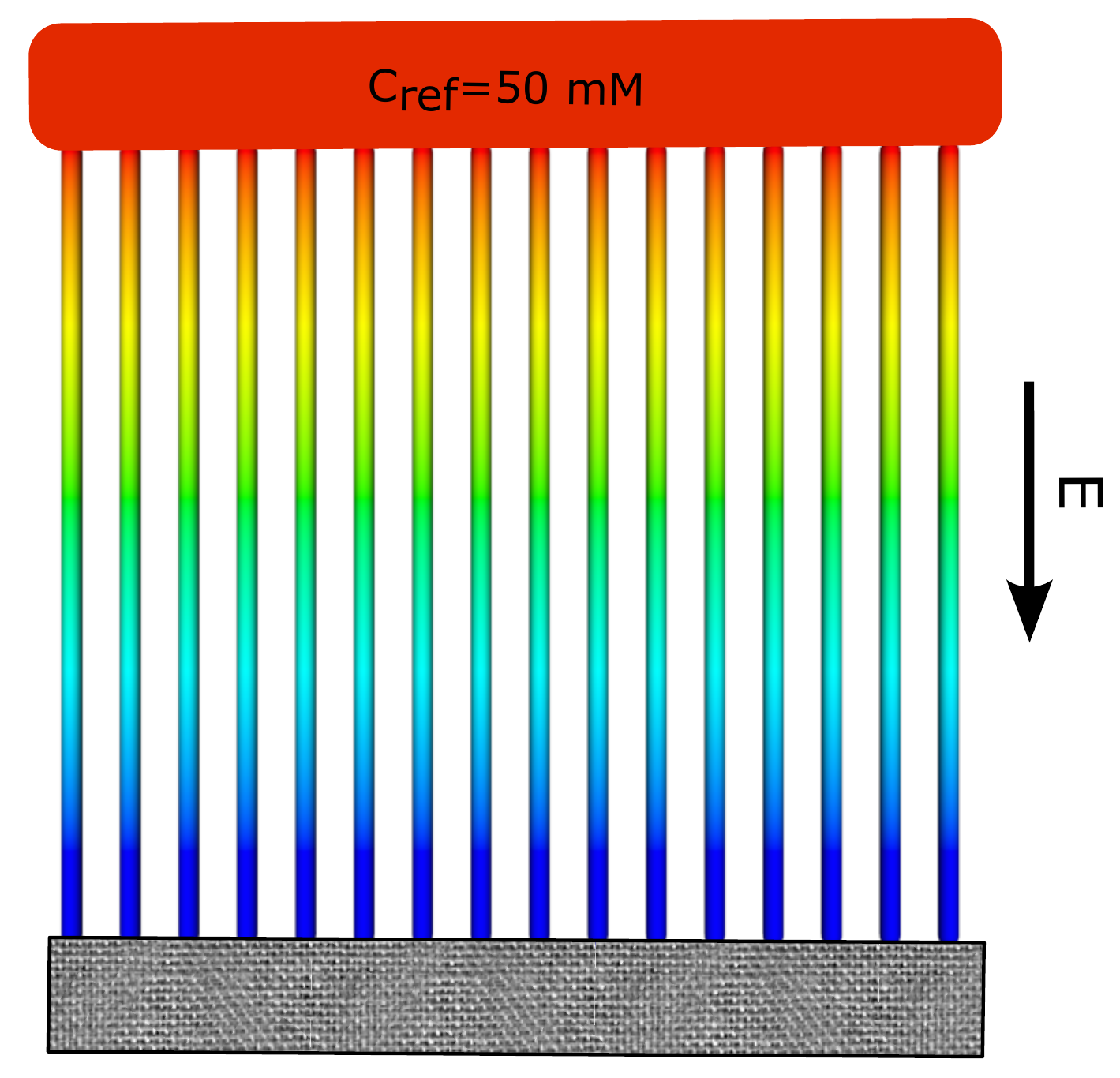}
                \caption{}
                \label{fig:homog_schematic}
        \end{subfigure}
        ~ 
         \caption{(a) The steady state concentration field in a hierarchical random network confined by a cation-selective membrane at the bottom and connected to a reservoir with uniform concentration of 50 mM at the top. (b) The concentration field of the homogenized model with the same effective pore size and $C_s$.}
 \label{fig:lattice_cfield}
 \end{figure}

\begin{figure}[H]
        \centering
        \begin{subfigure}[H]{0.4\textwidth}
                \includegraphics[width=\textwidth]{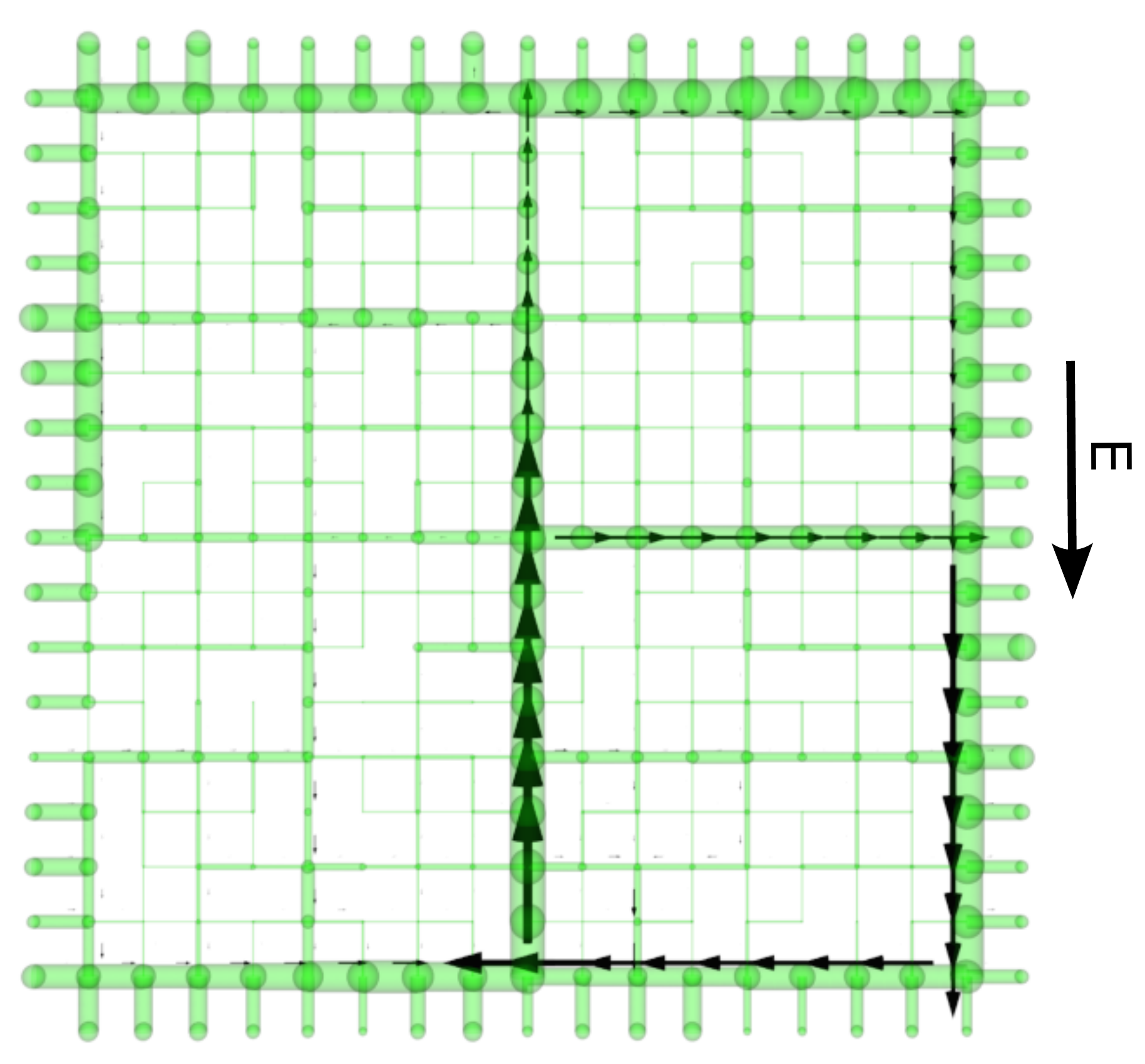}
                \caption{}
                \label{fig:flow}
        \end{subfigure}%
        ~
        \begin{subfigure}[H]{0.4\textwidth}
                \includegraphics[width=\textwidth]{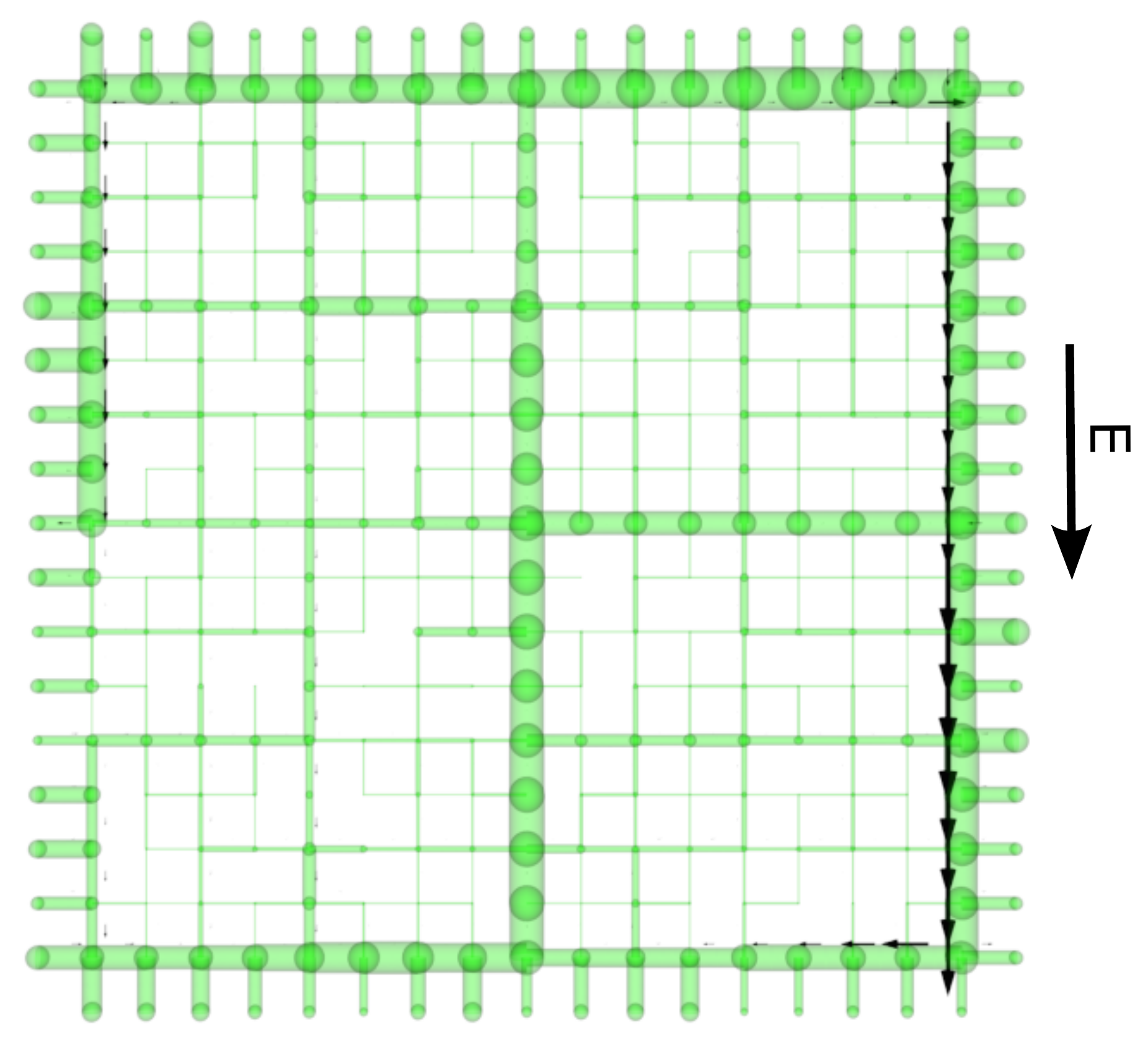}
                \caption{}
                \label{fig:curr}
        \end{subfigure}
         \caption{(a) The steady state flow field in the hierarchical random network. Despite the zero macroscopic flow across the system, there are induced flow loops with the length scales comparable to the length scale of the network. (b) The current field at steady state.  The full depletion of middle pores results in a high electric resistance that dramatically reduces the current flow in this portion of the domain. The right portion of the network which is affected by the electroosmotic flow from the reservoir to the membrane, contains higher bulk conductivity, which causes the majority of the charge transport to occur in this region.}
 \label{fig:lattice_flow_current}
 \end{figure}

We here present one sample simulation of a hierarchical network composed of four scales having moderate accessivity, indicating a combination of both parallel and series network connections between pores of different sizes. Using the methodology described in Section (\ref{sec:dimensionless parameters}), we obtained an intermediate accessivity is $\alpha \approx 0.114$ for this particular network.The mean size of the coarsest ``macropores'' is $2\mu m$, and finer ``micropores'' have proportionally smaller mean pore diameters at each level of recursion. The pore variability $C_v$ and pore blocking probability ($p$) are set equal to 0.3 and 0.2 respectively for all scales. The final network is depicted in Figure (\ref{fig:hetero_schematic}). It consists of 557 pores and 340 nodes. The top reservoir contains KCl solution with concentration of $C_{\text{ref}}=50 mM$ and the applied potential is $V/V_T=60$ at this boundary. The considered network has a uniformly distributed surface charge density equal to $q_s = -13 mC/m^2$. This setting results in the effective $h_p = 612.3 nm$, which corresponds to the effective $\lambda_D=0.002$, and the effective $C_s = 0.004$. Alongside the random network, we considered the equivalent homogenized model with the same $h_p$, and $C_s$. Similar to Section (\ref{sec:nonhierarchical structures}), we compare the steady state responses of the two systems including their concentration fields, flow fields and I-V characteristics. 
 Figure (\ref{fig:lattice_cfield}) indicates the steady state concentration fields for the random network and the homogenized model, when a potential gradient of  $V/V_T=60$ is applied. As opposed to the homogenized model, where the concentration field is 1D, the random network involves a heterogeneous pattern for the propagation of deionization shocks leading to a 2D and complex concentration field. Moreover, the depletion zone is located close to the membrane interface in the homogenized model, while there is a dominant deionization shock in the middle of the random network, which has propagated almost all the way to the top boundary. As shown in Figure (\ref{fig:flow}), this shock has propagated in the direction of the pressure-driven flow away from the membrane and has completely depleted the bulk concentration in the middle vertical pores. Additionally, the rightmost pores in the network involve a dominant electroosmotic flow towards the membrane, which impedes the propagation of the deionization shock.  
Figure (\ref{fig:flow}) and (\ref{fig:curr}) depict respectively the steady state flow field and the electric current field in the hierarchical network. As discussed before, although the macroscopic mean flow rate of the system is zero, there are internally induced flow loops resulting from the competition between electroosmotic flow and the backward pressure-driven flow. The internal flow loops would not occur if the pore sizes were uniform over the whole network, and having pore size variability is a key factor to provoke the competition of different modes of advection. As shown in Figure (\ref{fig:flow}), there are two major flow loops located on the right section of the network, which is the region containing pores with less hydraulic resistance than the bottom section of the medium. The strong backward pressure-driven flow in the middle vertical pore has been balanced by the dominant  electroosmotic flow on the right side of the network. The length scale associated with the largest loop is equal to the length scale of the system in the x direction. We obtained the Pe$=uL/D = \frac{\Sigma_{i=1}^n A_i u_i}{\Sigma_{i=1}^n A_i}\frac{L}{D}$ (where n is the number of pores forming the loop) associated with this loop to be 8.1, which demonstrates that the induced advection is remarkably dominant over the molecular diffusion in this porous structure.
 
Considering both diagrams in Figure (\ref{fig:lattice_flow_current}), one can see that the majority of the charge transport occurs in the rightmost vertical pores, where the fast flow is towards the membrane. The induced flow transports salt from the reservoir enhancing local electric conductance. Moreover, the current is almost zero in the middle vertical pathway as the backward pressure-driven flow has pushed the deionization shock towards the reservoir, leading to a substantial increase in the electric resistance. Since the system operates at low $C_s$, the charge transport due to surface conduction effect is small. 

\begin{figure}[H]
\makebox[\textwidth][c]{\includegraphics[width=0.5\textwidth]{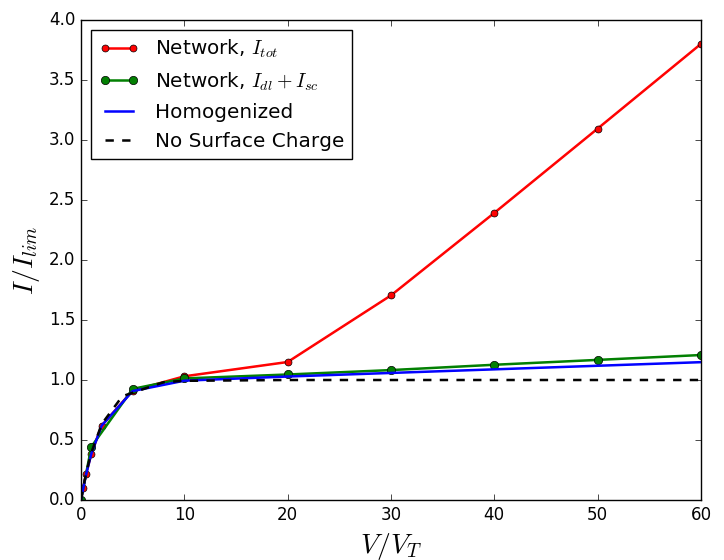}}
\caption{The I-V characteristics of the hierarchical network (red) with $p=0.2$, $C_v=0.33$, $\lambda_D=0.002$, $C_s=0.004$, and $\kappa=0.5$. The I-V curve of the same network with flow turned-off (green), I-V of the corresponding homogenized model (blue), and the diffusion-limited current (dashed black curve). Only a small portion of the current is transferred via surface conduction, and the majority is dominated by the advection mechanism.}
\label{fig:lattice_total_sc_current}
\end{figure}

The I-V characteristics of the hierarchical network and the homogenized model were obtained over a wide range of applied potential, as shown in Figure (\ref{fig:lattice_total_sc_current}). The homogenized model which is driven primarily by surface conduction has hardly moved above the limiting current, given that the system surface conduction is small. However, the hierarchical network has exhibited a significantly different response.  For $V/V_T<20$, the two systems show similar I-V responses. For intermediate voltages ($5<V/V_T<20$), the induced flow field is not strong enough and the system is mainly controlled by surface conduction. However, after $V/V_T=20$ the overlimiting conductance increases considerably in the hierarchical random network compared to the homogenized model. This voltage can be regarded as a transition point from which the induced flow loops dominates the weak surface conduction effect and allows a substantial charge transport towards the membrane. Our results demonstrate that after the transition point, the P{\' e}clet numbers associated with the flow loops of the system become higher than 1.

In Figure (\ref{fig:lattice_total_sc_current}), the green I-V curve denoted by $I_{dl}+I_{sc}$ belongs to the simulation of the hierarchical network where the advection terms are turned off. The I-V curve decomposition demonstrates that the surface conduction effect is well represented by the homogenized model, and it reconfirms the fact that all of the significant excess current in the full model must be due to the advection effect by the flow through the closed loops connecting multiple pores. This important advection mechanism is missing in the homogenized capillary bundle model.

We used our model to identify the dependency of overlimiting transport in hierarchical networks to $C_v$, $p$ and reported our investigation in Supplementary Information. Our results indicate that in all hierarchical networks there is a remarkable contribution of advection to overlimiting transport, and it is the dominant transport mechanism in the range of $C_s$ considered in this study. We consider hierarchical structures and aim at finding scaling laws relating the overlimiting conductance to system salinity, and examined the validity of theoretical scaling laws reported in the literature \cite{deng2013, dydek2011} in high and low surface conduction regime.

\subsection{Dependency on Bulk Salinity}
Following the experiments of Deng et al. \cite{deng2013}, we here present our investigation on the dependency of the I-V curve on system salinity in both low surface conduction and high surface conduction regime. Measurements of Deng et al. \cite{deng2013} matched the power law dependence of $\sigma_{olc}$ on $C_{\text{ref}}$ which was derived by Dydek et al. considering a single micro-channel\cite{dydek2011}:
\begin{equation}
\tilde{\sigma}_{\text{olc,adv}} \sim q_s^{2/5} C_{\text{ref}}^{4/5}. 
\label{eq:dydek_scaling}
\end{equation}

Our analysis presented in Section (S2.1) in Supplementary Information, as well as the scaling estimates based on the theory of Dydek et al \cite{dydek2011}, suggests that a single-pore circulation mechanism cannot result in significant  $\sigma_{\text{olc,adv}}$ in the regimes considered in the experiment. Given the results presented in earlier sections, only hierarchical structures may sustain significant $\sigma_{olc,adv}$. Therefore, in this section we investigate the scaling of the overlimiting conductance with respect to reservoir concentration ($C_{ref}$) for only hierarchical porous networks. 

In our study we maintained $\kappa=0.5$,  $p=0.2$, $C_v=33$, and $C_s$ fixed as we vary $C_{\text{ref}}$. This is equivalent to only changing $\lambda_D$. We intentionally chose to maintain fixed $C_s$ (instead of fixed $q_s$) since change of $C_s$ can transition the system from being dominated by surface conduction ($\sigma_{\text{olc}} \sim\sigma_{\text{olc,sc}}$) to that dominated by advection ($\sigma_{\text{olc}}\sim \sigma_{\text{olc,adv}}$). As we shall see, different scalings with respect to $C_{\text{ref}}$ are obtained in high and low surface conduction regimes. As constant $C_s$ implies $\sigma \sim C_{\text{ref}}$, the scaling proposed by Dydek et al. \cite{dydek2011} in equation (\ref{eq:dydek_scaling}) suggests a 6/5 scaling of $\sigma_{\text{olc,adv}}$ with respect to $C_{\text{ref}}$. 

Note that to eliminate the effect of statistical variations, we require to keep the network geometry fixed, as commonly done in experiments. As will be discussed, in high $C_s$ regime, the overlimiting conductance is primarily driven by surface effect ($\sigma_{\text{olc}} \simeq \sigma_{\text{olc, sc}}$). Conversely, in the limit of low surface conduction, the advection mechanism becomes dominant, and the charge transport is controlled primarily by the flow loops in the network ($\sigma_{\text{olc}} \simeq \sigma_{\text{olc, adv}}$).

\subsubsection{High Surface Conduction Regime}
We considered different $C_s$ values in the limit of high surface conduction regime, and computed the I-V characteristics over a range of $C_{ref}$. The hierarchical network introduced in Section (\ref{sec:sample_hierarach_results}) was considered for this study. To keep $C_s$ fixed, one needs to increase the surface charge of the network proportionally as $C_{ref}$ increases. Our results demonstrate that in the high surface conduction limit the contribution of surface conduction to charge transport always dominates advection mechanism, leading to $\Gamma<1$.  
\begin{figure}[H]
        \centering
        \begin{subfigure}[H]{0.44\textwidth}
                \includegraphics[width=\textwidth]{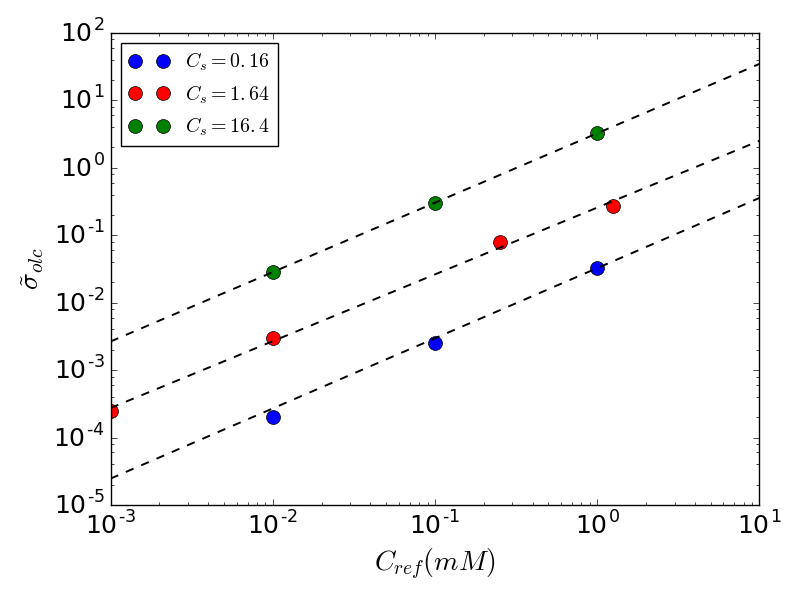}
                \caption{}
                \label{fig:olc_cref_high}
        \end{subfigure}%
        ~ 
        \begin{subfigure}[H]{0.44\textwidth}
                \includegraphics[width=\textwidth]{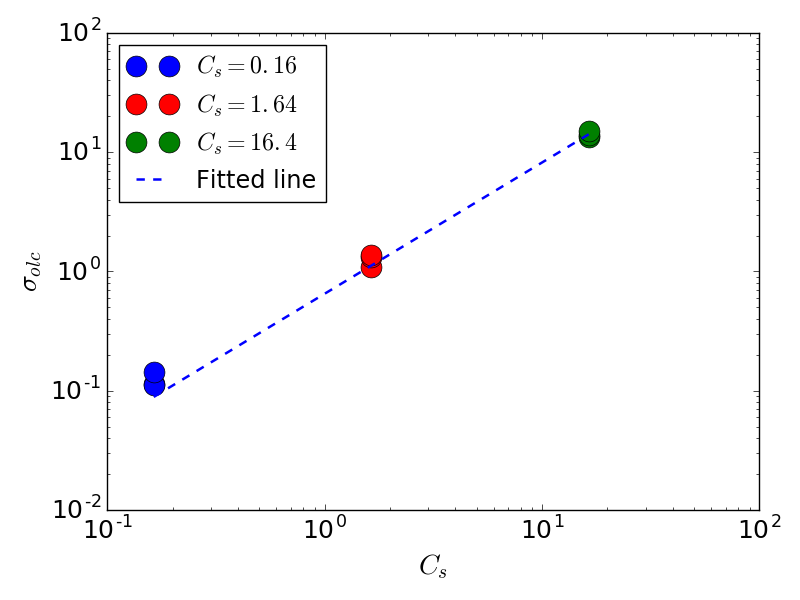}
                \caption{}
                \label{fig:olc_cs_high}
        \end{subfigure}
        ~ 
         \caption{(a) The dimensional overlimiting conductance ($\tilde{\sigma}_{\text{olc}}$) vs. $C_{\text{ref}}$ for three values of $C_s$ in the high surface conduction limit. All three cases reveal linear scaling of the conductance with respect to $C_{\text{ref}}$ in the overlimiting regime. (b) The dimensionless overlimiting conductance ($\sigma_{\text{olc}}$) vs. $C_s$ for $C_{\text{ref}}=1mM$. A linear relation is observed between surface conduction and overlimiting conductance, which is consistent with theoretical results \cite{deng2013, dydek2011}.}
 \label{fig:two_lattice}
 \end{figure}

We have plotted the dimensional overlimiting conductance ($\tilde{\sigma}_{olc}$) versus $C_{\text{ref}}$ for three values of $C_s$ in Figure (\ref{fig:olc_cref_high}). This result reveals that for high surface conduction regime, the overlimiting conductance is linearly proportional to $C_{ref}$. Similarly, we observed the linear relation between the overlimiting conductance and $C_s$, which is shown in Figure (\ref{fig:olc_cs_high}). Using the homogenized model in the limit of thin double layers, one can use the area-averaged transport equation in high surface conduction regime to demonstrate the linear scaling theoretically as follows \cite{dydek2011}:
\begin{equation}
\frac{I}{I_{\text{lim}}} = 1 - \exp^{-\Delta V /V_T} + C_s \Delta V/V_T,
\label{eq:olc_sigma_cs}
\end{equation}
where $\Delta V$ refers to the applied electric potential difference between the reservoir and the membrane. In the limit of high $\Delta V$, the overlimiting conductance is  
\begin{equation}
\sigma_{\text{olc}} = \frac{dI/I_{\text{lim}}}{d \Delta V/V_T}= C_s. 
\label{eq:sigma_olc}
\end{equation}

Note that although we compute $\sigma_{\text{olc}}$ with respect to cation electrochemical potential difference (we use $V/V_T = \Delta \mu^+$ instead of $\Delta V/V_T$ as the input), the relation in (\ref{eq:sigma_olc}) still holds for our numerical results. This is because most of the electric potential gradient occurs in the depleted region, where $\bar{C}^+$ is almost constant, $ \bar{C}^+\simeq C_s$, and the cation electrochemical potential difference is primarily due to electric potential gradient, and hence, $\Delta \mu^+ \simeq \Delta V/V_T$.

Given $I_{\text{lim}}$ is the outcome of a solution to the linear Laplace equation, it must be proportional to $C_{\text{ref}}$ regardless of the network geometry. Therefore, from equation (\ref{eq:sigma_olc}), it is expected that $\tilde{\sigma}_{\text{olc}}$ which is equal to $\sigma_{\text{olc}} \times I_{\text{lim}}/V_T$, grows proportional to $C_{\text{ref}}$ for a fixed $C_s$. Additionally, when $C_{\text{ref}}$ is fixed, this expression predicts the proportionality of $\sigma_{\text{olc}}$ with $C_s$, as observed in Figure (\ref{fig:olc_cs_high}). These observations further confirm the fact that in the high $C_s$ limit advection plays negligible role in manipulating the I-V response.
 
\subsubsection{Low Surface Conduction Regime}
We here employed our model to simulate the electrokinetic transport in hierarchical networks for different values of $C_{\text{ref}}$ while fixing $C_s$ to 0.004. In Section (\ref{sec:sample_hierarach_results}) we showed that for a low value of $C_s$ advection could be the dominant mechanism for overlimiting transport in this configuration. We considered a wide range of $C_{\text{ref}}$ and investigated whether there is a power law relating the overlimiting conductance to $C_{\text{ref}}$, as explored in \cite{deng2013}. To maintain a small $C_s=0.004$, the surface charge density of the system is proportionally decreased as we decreased $C_{\text{ref}}$.

\begin{figure}[H]
        \centering
        \begin{subfigure}[H]{0.4\textwidth}
                \includegraphics[width=\textwidth]{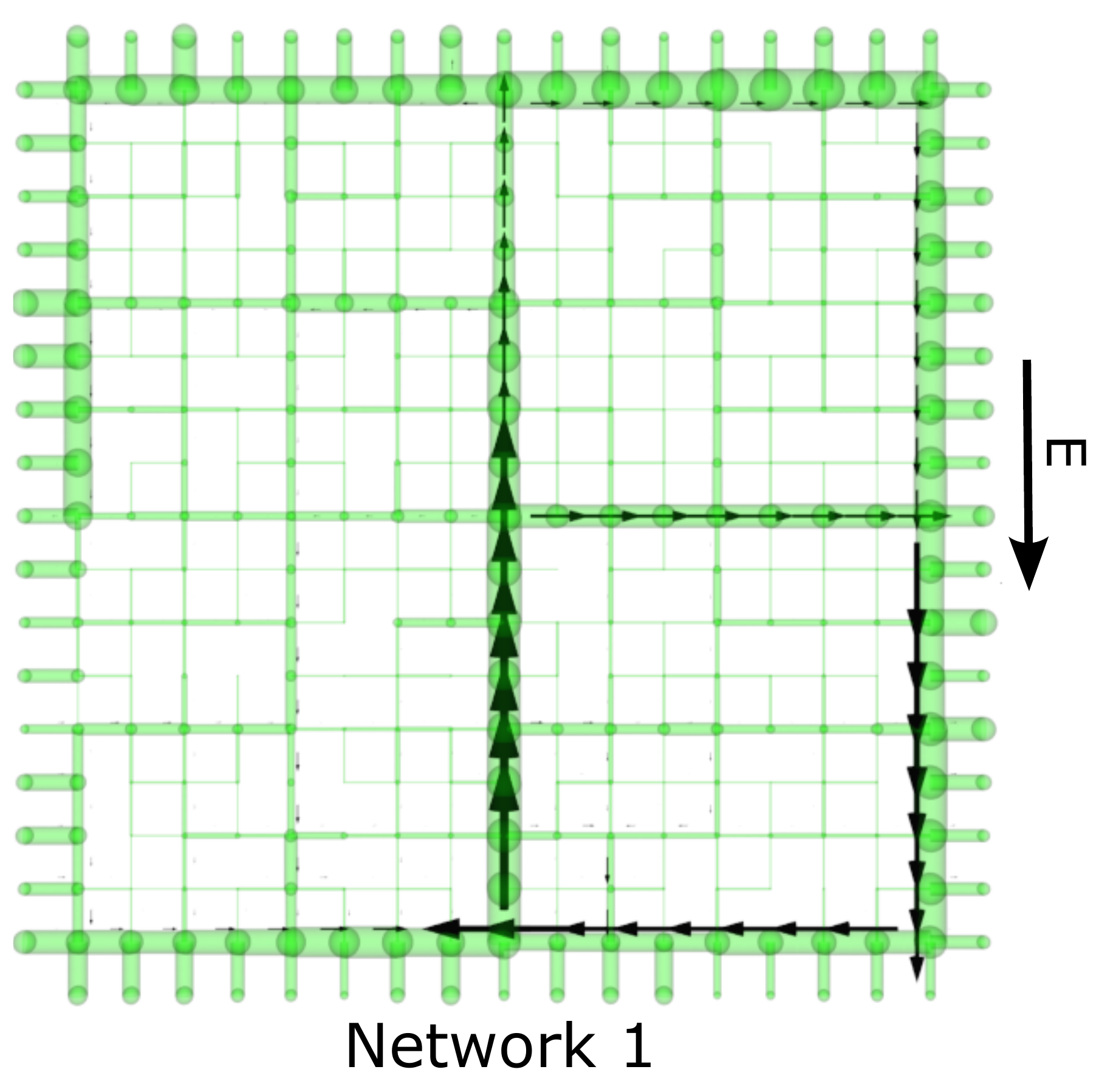}
                \caption{}
                \label{fig:hetero_s42}
        \end{subfigure}%
        ~ 
        \begin{subfigure}[H]{0.4\textwidth}
                \includegraphics[width=\textwidth]{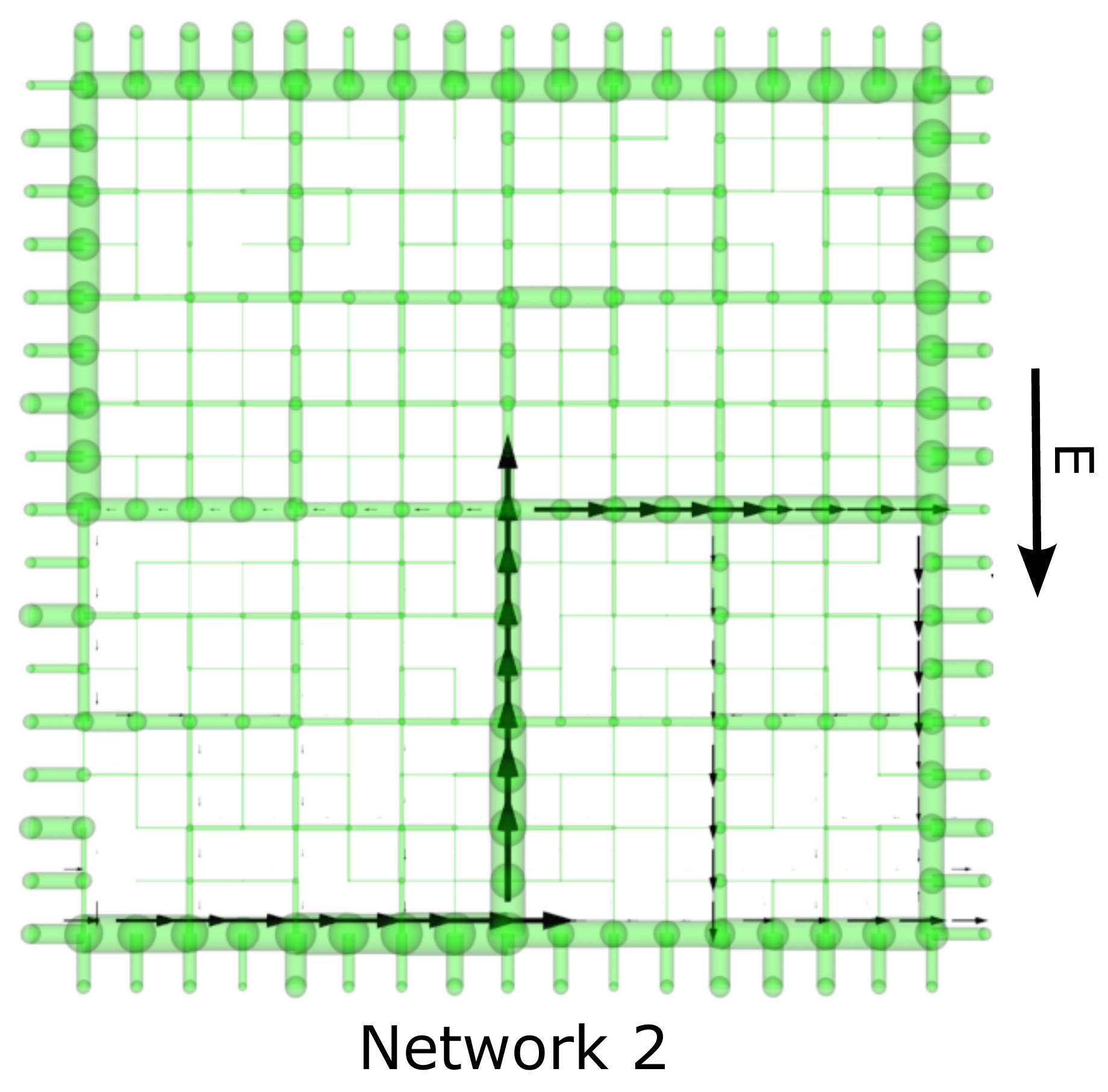}
                \caption{}
                \label{fig:hetero_s1897}
        \end{subfigure}
        ~ 
         \caption{Two hierarchical networks and their internally induced flow fields when $V/V_T=50$ is applied across the systems. Network 1 which is shown in (a) corresponds to the network whose response was discussed comprehensively in Section (\ref{sec:sample_hierarach_results}). The width of the flow arrows is proportional to the magnitude of the flow in that location. Network 1 involves large scale flow loops that act faster than the induced flow loops in Network 2 shown in (b).}
 \label{fig:two_lattice_geom}
 \end{figure}

\begin{figure}[H]
\makebox[\textwidth][c]{\includegraphics[width=0.5\textwidth]{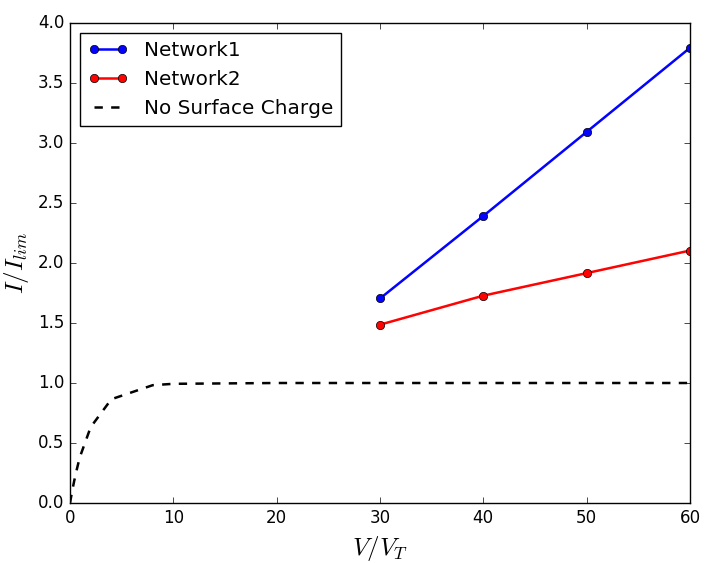}}
\caption{I-V characteristics of the two hierarchical networks shown in Figure (\ref{fig:two_lattice_geom}) in the overlimiting regime. Higher charge transport occurs in Network 1 which involves faster and larger induced flow loops compared to Network 2.}
\label{fig:iv_two_lattice}
\end{figure}

We present our results for two hierarchical porous networks shown in Figure (\ref{fig:two_lattice_geom}). Both porous structures are generated using the same input parameters, $C_v=0.33$, and $p=0.2$, $C_s=0.004$, and $\kappa=0.5$. For the given input parameters, we sampled different sets of random variables for blocking pores as well as the pore diameters. The effective $h_p$ in these structures are very close and equal to 612 nm. 
When the reservoir concentration is $C_{\text{ref}}=50mM$, the internally induced flow fields at $V/V_T=50$ are shown in Figure (\ref{fig:two_lattice_geom}) demonstrating that the strength and the length scales of the induced flow loops are different due to the randomness in the network geometries. We have compared the I-V response of the two porous systems obtained for $C_{\text{ref}}=50 mM$ in Figure (\ref{fig:iv_two_lattice}). The majority of the charge transport is driven by advection mechanism in both systems as the surface conduction effect is small. Yet, the advection effect is more dominant in Network 1 leading to the transport of higher amount of charge through this network. 

\begin{figure}[H]
        \centering
        \begin{subfigure}[H]{0.4\textwidth}
                \includegraphics[width=\textwidth]{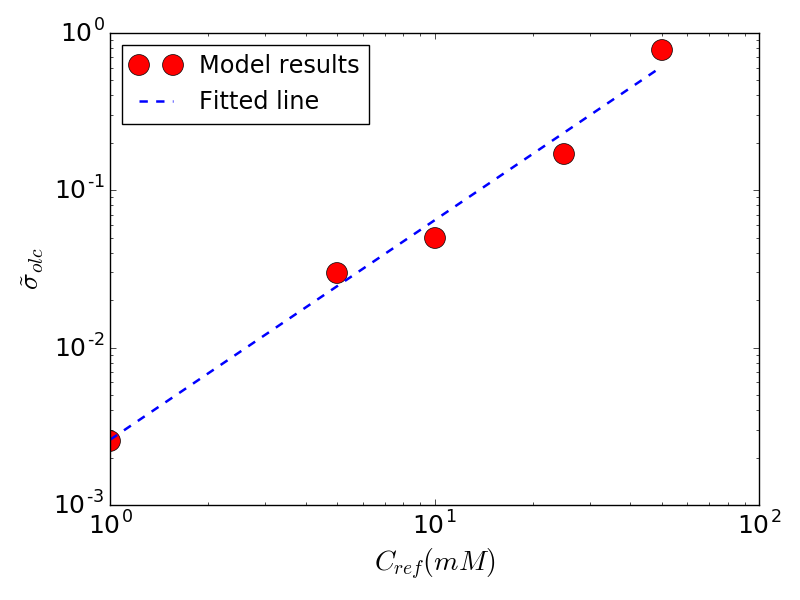}
                \caption{}
                \label{fig:olc_cref_low1}
        \end{subfigure}%
        ~ 
        \begin{subfigure}[H]{0.4\textwidth}
                \includegraphics[width=\textwidth]{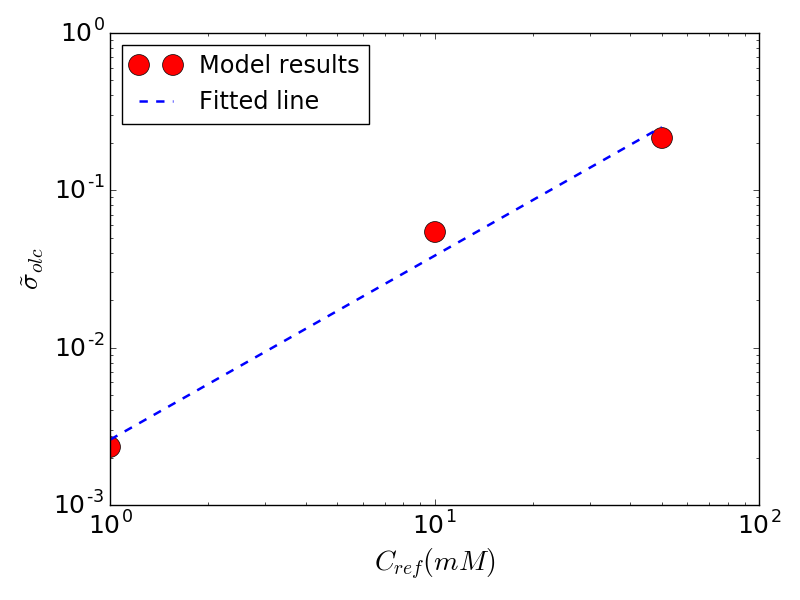}
                \caption{}
                \label{fig:u_cref_low2}
        \end{subfigure}
        ~ 
         \caption{(a) The dimensional overlimtiing conductance vs. $C_{\text{ref}}$ in log-scale for Network 1 (a) and Network 2 (b). For low to moderate values of concentration fitted scaling power of 1.4 and 1.2 are obtained respectively for Network 1 and Network 2.}
 \label{fig:olc_two_lattice}
 \end{figure}

Figure (\ref{fig:olc_two_lattice}) depicts the dimensional overlimiting conductance versus $C_{\text{ref}}$ in the log-scale for two hierarchical networks. For low to moderate values of $C_{\text{ref}}$ we obtained the 1.4 scaling of $\tilde{\sigma}_{olc}$ with respect to $C_{\text{ref}}$ for Network 1, whereas the scaling associated with Network 2 turns out to be 1.2. Furthermore, the overlimiting conductance deviates to lower values for $C_{\text{ref}}$ larger than 50 mM, and the scaling does not hold.
Here we sought a power law scaling following the experimental work by Deng et al. \cite{deng2013}. However, the plots shown in Figure (\ref{fig:olc_two_lattice}) suggest that power laws are not the best fit to the observed trends. Furthermore, even the fitted power depends on the specific geometry, and thus is not universal. Interestingly, for one of the networks the fitted power law is equal to 1.2, which matches the theoretical prediction of Deng et al. \cite{dydek2011, deng2013}. In this work, however, demonstrate that a power law with exponent 1.2 (or any other value) cannot be universal for all types of porous networks.  Since the glass frit of used by Deng {\it et al.} consists of fused and sintered silica particles, the porous network was likely not very hierarchical, so it may also  be that additional mechanisms affected the results, such as charge regulation of the surfaces, local pH variations, etc. 

\begin{figure}[H]
        \centering
        \begin{subfigure}[h]{0.44\textwidth}
                \includegraphics[width=\textwidth]{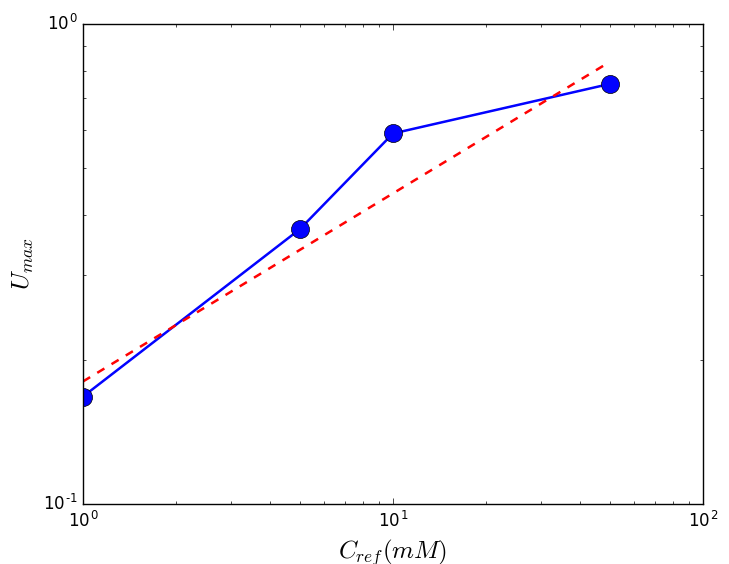}
                \caption{}
                \label{fig:u_cref_low1}
        \end{subfigure}%
        ~ 
        \begin{subfigure}[h]{0.44\textwidth}
                \includegraphics[width=\textwidth]{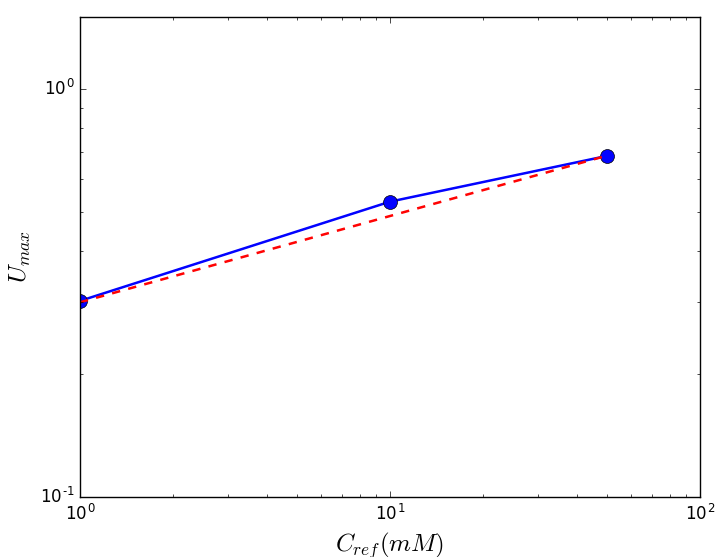}
                \caption{}
                \label{fig:u_cref_low2}
        \end{subfigure}
          \begin{subfigure}[h]{0.5\textwidth}
                \includegraphics[width=\textwidth]{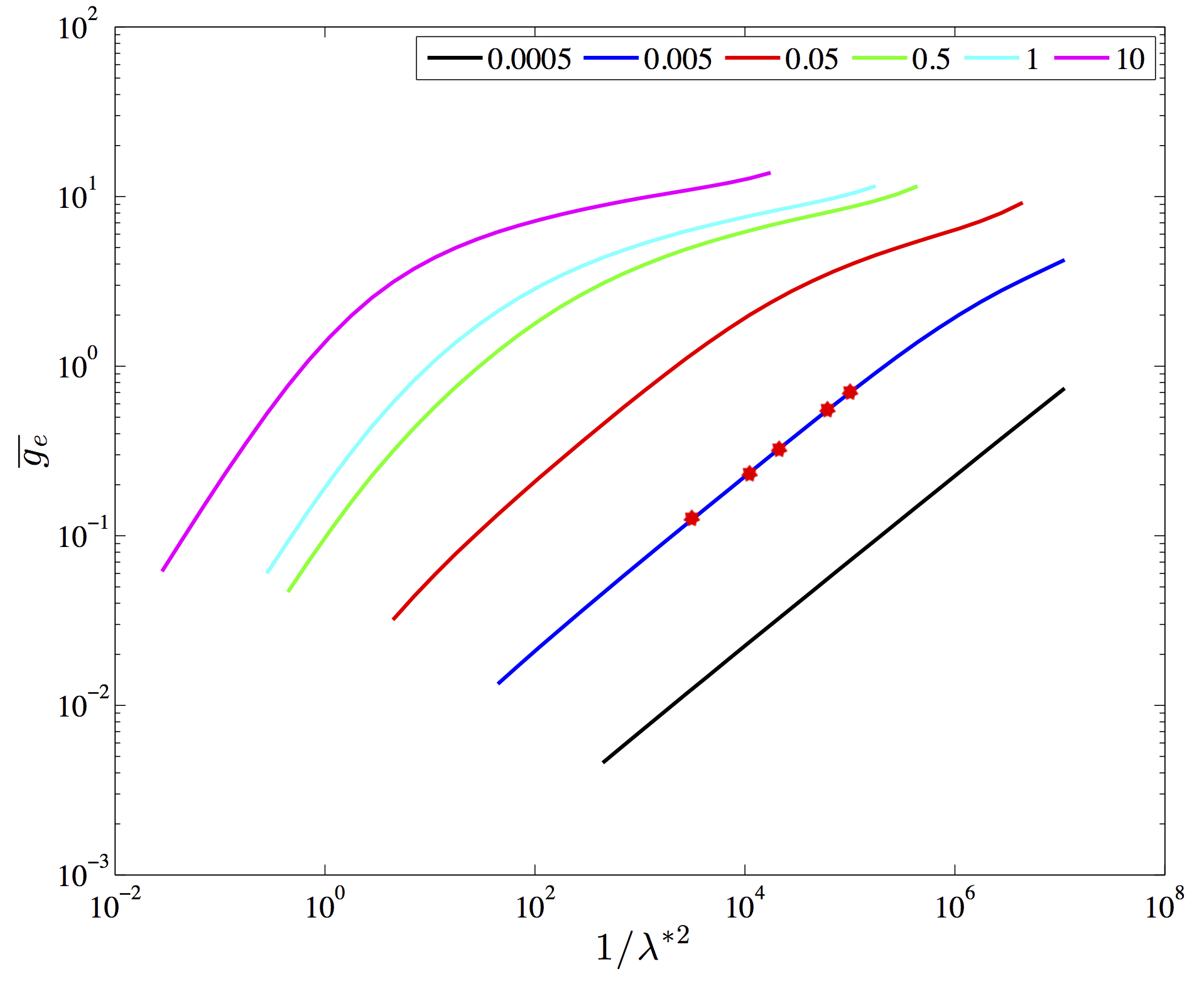}
                \caption{}
                \label{fig:ge_cref}
        \end{subfigure}
         \caption{The variation of dimensionless maximum velocity in Network 1 (a) and Network 2 (b) with respect to $C_{\text{ref}}$ in log-scale. Our numerical results show that for Network 1, $U_{\text{max}} \sim C_{\text{ref}}^{0.4}$ and for Network 2, $U_{\text{max}} \sim C_{\text{ref}}^{0.2}$. (c) The variation of the electroosmotic velocity prefactor ($\bar{g}_e$) vs. $1/\lambda_D^2$ in the log-scale for different values of $C_s$. At low surface conduction limit, the plot indicates 1/2 scaling. The curves deviate from this scaling for large values of $1/\lambda_D^2$ (large concentration). Moreover, increasing the surface conduction results in different scaling laws for $\bar{g}_e$, implying that the electroosmotic flow cannot be predicted by a single scaling law.}
 \label{fig:u_ge_cref}
 \end{figure}

Next, we provide deeper investigation on the trends of $\sigma_{\text{olc,adv}}$ obtained from our simulations. Before doing so, we provide some context leading to a plausible explanation for the observed behavior. By analogy to the scaling of $I_{\text{lim}}\sim zeDC_{\text{ref}}/L_x$ where $L_x$ is the length scale of the network, consider $I_{\text{olc}}\sim zeD_{\text{eff}}C_{\text{ref}}/L_x$ for overlimiting regime, where the effective diffusivity $D_{\text{eff}}$ incorporates the additional dispersion due to induced flow loops. In the overlimiting regime with dominant advection mechanism, $D_{\text{eff}}$ can be viewed as eddy diffusivity which scales as $U_{\text{eddy}}l_{\text{eddy}}$. As demonstrated in Section (\ref{sec:sample_hierarach_results}), in the overlimiting regime the length scale of the induced flow loop is comparable to the network length scale ($l_{\text{eddy}} \sim L_x$). Using maximum magnitude of velocity ($U_{\text{max}}$) for the scaling of $U_{\text{eddy}}$ yields $I_{olc}\sim zeD_{\text{eff}}C_{\text{ref}}L_x \sim U_{\text{max}} C_{\text{ref}} $. We investigated the relation between $U_{\text{max}}$ and $C_{\text{ref}}$ (Figure (\ref{fig:u_ge_cref})) and observed the power fits with the exponents 0.4 and 0.2 respectively for Networks 1 and 2. This yields $I_{\text{olc}}\sim C_{\text{ref}}^{1.4}$ for Network 1 and $I_{\text{olc}}\sim C_{\text{ref}}^{1.2}$ for Network 2, which are close to the fitted results of $\sigma_{\text{olc}}$ obtained for low and moderate values of $C_{\text{ref}}$. This observation could imply that the flow loops with high Pe number are the key controlling factor of the network behavior in the low surface conduction regime.
Lastly, we provide insights on why  $U_{\text{max}}$ may increase with the electrolyte salinity when $C_s$ is held fixed. Using our tabulated area-averaged coefficients \cite{alizadeh1}, we re-plotted $\bar{g}_e$ coefficient, representing the electroosmotic velocity, with respect to $1/\lambda_D^2\sim C_{\text{ref}}$ for different values of $C_s$ (Figure (\ref{fig:ge_cref})). This diagram illustrates that for a fixed $C_s$ increasing $C_{\text{ref}}$ results in faster electroosmotic flow in a single channel subject to a fixed electric potential. In the thin EDL limit, this trend is expected given that channel zeta potential ($\zeta$) is increased as both $C_{\text{ref}}$ and $q_s$ are proportionally increased.  However, moving from low surface conduction to high surface conduction regime can result in various scaling of electroosmotic flow even in a single pore. 
 The largest slope at any point on the curves for low $C_s$ indicates a 1/2 power law for the dependency of $U_{\text{max}}$ on $C_{\text{ref}}$, which is higher than the power law fits obtained for the two networks in Figures (\ref{fig:u_cref_low1}) and (\ref{fig:u_cref_low2}). The marked points on Figure (\ref{fig:ge_cref}) indicate the regime examined in the considered random networks. Coincidentally, these points correspond to the maximum scaling slope. Therefore, the lower power exponent, obtained in the data fit from the measurement of $U_{\text{max}}$ can be attributed to the pore coupling effects and partial balance of the electroosmotic velocity by pressure-driven flows. 

 Aside from complexities arising from pore coupling and specific network geometry, figure (\ref{fig:ge_cref}) suggests that even in the simplest geometries there is no universal power law for the dependence of advective effects on $C_{\text{ref}}$. As $C_{\text{ref}}$ varies, the transition from thin to thick EDL and low to high zeta potential can significantly shift the trends in the induced electroosmotic flow. This complexity as well as aforementioned sensitivity to geometric details of a network, brings significant challenges to the task of characterization of transport in porous media. 
 
\section{Correlation with Accessivity}
We investigated the correlation of accessivity and $\sigma_{OLC}$ for different network topologies considered in this study. For non-hierarchical networks by computing the ratio of number of pores connected to membrane surface to the total number of pores, we obtained low accessivity $\alpha \simeq 0.03$. For hierarchical networks as discussed in previous sections, we utilized a slightly different method to compute accessivity and obtained moderate accessivity $\alpha \simeq 0.12$ for hierarchical lattices constructed for different network parameters $C_v$ and $p$.

Porous structures that exhibit strong EO eddies around network loops tend to have intermediate values of accessivity, far from the limits of predominantly parallel ($\alpha=1$) and series ($\alpha=0$) connections between pores of different sizes.  Networks with intermediate values of accessivity and intricate combinations of parallel and series couplings also tend to be {\it isotropic}, with connected paths of accessible pores in at least two directions, thus allowing for many loops in the network. In contrast, the homogenized capillary-bundle model has the maximum accessivity ($\alpha=1$) for surfaces perpendicular to current direction (such as the bounding membrane), but the minimum accessivity ($\alpha=0$) for restricted to surfaces in the transverse directions, thus eliminating the possibility of any flow loops.   Our hierarchical networks that exhibit massive increases in OLC from EO convection loops are isotropic with moderate values of accessivity to bounding surfaces, via connected pathways of macropores, in both directions.

\begin{figure}[H]
\makebox[\textwidth][c]{\includegraphics[width=0.6\textwidth]{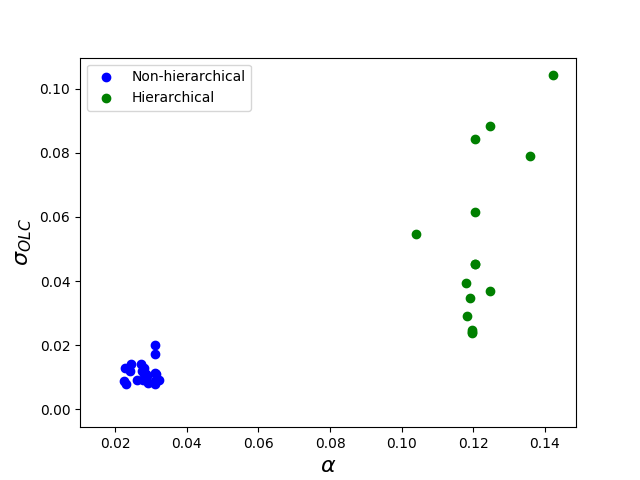}}
\caption{Nondimensional overlimiting conductance versus accessivity for the wide variety of non-hierarchical and hierarchical networks considered in this study. $C_{\text{ref}}=50mM$ is used to compute $\sigma_{\text{OLC}}$ in all porous networks. }
\label{fig:sigma_alpha}
\end{figure}

Figure(\ref{fig:sigma_alpha}) depicts the variation of  $\sigma_{OLC}$ with respect to accessivity for various networks that we investigated in this article. Our data have formed two main clusters and are colored based on the type of network topology. The results associated with non-hierarchical networks are depicted by blue points and the results obtained for hierarchical networks are shown in green. All data points associated with non-hierarchical networks are placed in the left bottom corner of the graph, where represents low values of $\sigma_{OLC}$ and low values accessivity. These networks cannot create the macroscopic flow loops that are required to sustain larger $\sigma_{OLC}$. 
Hierarchical networks have resulted in intermediate values of accessivity ($\alpha \simeq 0.1$), but their $\sigma_{OLC}$ covers relatively a wide range. This implies that to have large convective over-limiting conductance, it is necessary to have combinations of parallel and series couplings in loops (intermediate accessivity), which corresponds to moderate values of accessivity. However, this is not the sufficient condition for having a large overlimiting conductance, and other parameters such as pore size distribution are required along with accessivity to precisely characterize nonlinear elektrokinetic phenomena in porous media.

\section{Conclusions}
In summary, we presented a comprehensive numerical study of electrokinetic transport in hierarchical and non-hierarchical networks of pores with random pore coupling and pore blocking. Our simulations reveal that pore random size and pores' coupling are necessary conditions to provoke internal flow loops that can considerably enhance charge transport through the porous system. When the P{\' e}clet number associated with these flow loops becomes larger than 1, the induced advection can lead to significant increase in the overlimiting transport and system performance. We demonstrated that this effect is completely missed by the homogenized model, leading to the mis-prediction of the physics in the porous structures. Our results suggest that the nonlinear electrokinetic characteristics of porous structures can be exploited for designing efficient and compact water deionization devices, separation techniques, and high performance micro-pumps. 

Using our computational framework, we investigated the dependency of the overlimiting conductance on different parameters including network  accessivity, pore size variation, pore blocking probability, and electrolyte salinity. We observed that for non-hierarchical networks of low accessivity, having mostly series connections from the membrane through pores of different sizes, the role of advection in overlimiting transport is not significant, even for fairly low $C_s$ value of $\sim0.01$. On the contrary, in hierarchical structures of moderate and isotropic accessivity, having many different-sized internal pores connected in parallel to the transport surfaces, the network advection mechanism can have a tremendous impact on overlimiting ionic transport. We demonstrated that there is no universal scaling on the dependence of $\sigma_{\text{olc, adv}}$ on $C_{\text{ref}}$, at least in finite networks considered here, since having a dominant advection regime in the low surface connection limit inherently depends on the network geometry. While this makes characterization a challenging task, our work suggests that there are many opportunities for geometrical optimization and robust design of porous materials by predicting the physical trends relating the system performance to network's geometrical parameters.  

\section{Acknowledgments}
The authors gratefully acknowledge the fruitful comments by Zongyu (Joey) Gu on the definition and characterization of accessivity in porous media. This work is supported by the National Science Foundation (NSF) under award no. 1553275.
\begin{suppinfo}
Derivation of Diffusion-limited Current. 
Non-hierarchical Networks: Dependency on Network Parameters. 
Hierarchical Networks: Dependency on Network Parameters. 
\end{suppinfo}

\bibliography{manuscript}


\end{document}




\section{Derivation of Diffusion-limited Current} \label{app:derivation of difflimit current}
\begin{figure}[H]
\makebox[\textwidth][c]{\includegraphics[width=0.6\textwidth]{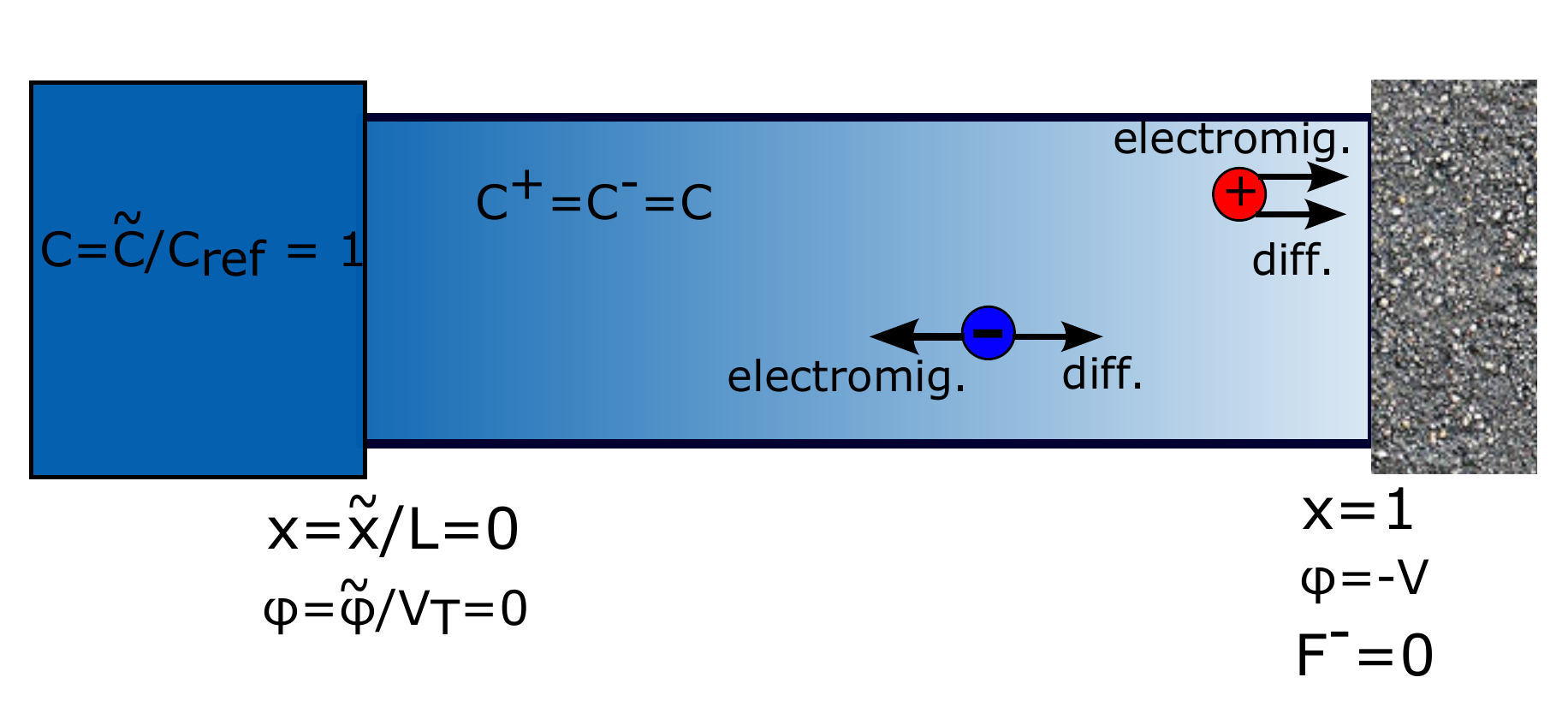}}
\caption{Schematic of a single pore with zero surface charge confined by an ideal cation-selective membrane.}
\label{fig:pore_difflimit}
\end{figure}
Consider a single pore with zero surface charge, subject to an external electric field. The field pushes the cations towards a cation-selective membrane as shown in Figure (\ref{fig:pore_difflimit}).  Concentration polarization leads to emergence of gradients in the salinity in the bulk fluid. Therefore, bulk concentration near the membrane drops to $c_b<1$, which is an unknown value to be determined. Due to electroneutrality in the pore, $C^+ = C^- = C$ along the pore. Moreover,  due to the cation selective membrane, the flux of cations are constant and equal to the system current density, and the flux of anions are zero, 
\begin{equation}
F^+ = -\frac{dC}{dx} - C\frac{d\phi}{dx} = i,
\label{eq:cation_flux_difflimit}
\end{equation}

\begin{equation}
F^- = -\frac{dC}{dx} + C\frac{d\phi}{dx} = 0.
\label{eq:anion_flux_difflimit}
\end{equation}
Equation (\ref{eq:anion_flux_difflimit}) yields $C=\exp(\phi)$ which results in $c_b=\exp(-V)$ at x=1. Moreover, the steady state concentration profile is governed by the diffusion equation which results in $C = 1 - (1-c_b)x$. Using equation (\ref{eq:cation_flux_difflimit}) we obtain the system current density,
\begin{equation}
i = -\frac{dC}{dx} - C\frac{d\phi}{dx} = -2\frac{dC}{dx} = 2(1-c_b) = 2(1-\exp(-V)).
\end{equation}
In dimensional format, $i = \frac{\tilde{i}}{zeD C_{\text{ref}}/L}$ and $V=\tilde{V}/{V_T}$. Thus,
\begin{equation}
\tilde{i} = i_{\text{lim}} (1-\exp(-\tilde{V}/V_T)),
\end{equation}
where, $i_{\text{lim}}  = \frac{2zeDC_{\text{ref}}}{L}$ is known as limiting current, which occurs when $\tilde{V} \gg V_T$. Considering the electrochemical potential of cations, $\mu^+ = \ln(C) + \phi$, one can obtain the gradient of electrochemical potential across the pore as follows,
\begin{equation}
\Delta \mu^+ = \mu^+(x=0) - \mu^+(x=1) = -\ln(c_b) + V = 2V.
\end{equation}
Thus, in the nondimensionalized format, the diffusion-limited current based on electrochemical potential is:
\begin{equation}
i = 1-\exp(-\Delta \mu^+/2).
\end{equation}
In the diffusion limited regime for large potential gradients (large electrochemical potentials), system current density exponentially converges to limiting current.

\newpage
\section{Non-hierarchical Networks: Dependency on Network Parameters} \label{sec:smr_p_dependency_nonhierarchical}
\subsection{Pore size variability}
To explore the impact of pore size variation, we change STD to mean ratio of pore sizes, $C_v$, in the random network, while holding the mean pore diameter and the rest of the parameters fixed. For all simulations, we considered mean pore diameter of 600 nm, uniform surface charge density $\sigma=-10 mC/m^2$, reservoir concentration of $C_{\text{ref}}=50 mM$, and pore blocking probability $p=0.35$. We maintain the topology of all network geometries by using the same pore blocking pattern, while changing the pore sizes by sampling different sets of pore diameters for a given mean and $C_v$ value. We ran simulations for $C_v=\{ 0.0, 0.15, 0.33, 0.66\}$. This led to effective pore sizes respectively equal to $h_p=\{150, 152, 170, 220\}$ nm. The effective Debye lengths associated with the networks, $\lambda_D$, are respectively $\{0.009, 0.009, 0.008, 0.006\}$. Given these parameters, all of the simulations were conducted in the small surface conduction regime to enable the assessment of the contribution of advection mechanism to ion transport. Effective $C_s$ for the reported $C_v$ values in this study are respectively $C_s=\{0.013, 0.013, 0.012, 0.009\}$. Our numerical simulations of nonhierarchical networks indicate that for fixed $p$ and variable $C_v$, the results are less sensitive to statistical variation than when $p$ varies and $C_v$ is held fixed. Nonetheless, we considered two simulations of large nonhierarchical networks (shown in Figure (1a)) for each set of network parameters using two separate sets of random numbers, and report the ensemble averaged results. The statistical variation of normalized currents, $I/I_{\text{lim}}$, is about 3\% of the reported averaged quantities.

\begin{figure}[H]
\makebox[\textwidth][c]{\includegraphics[width=0.55\textwidth]{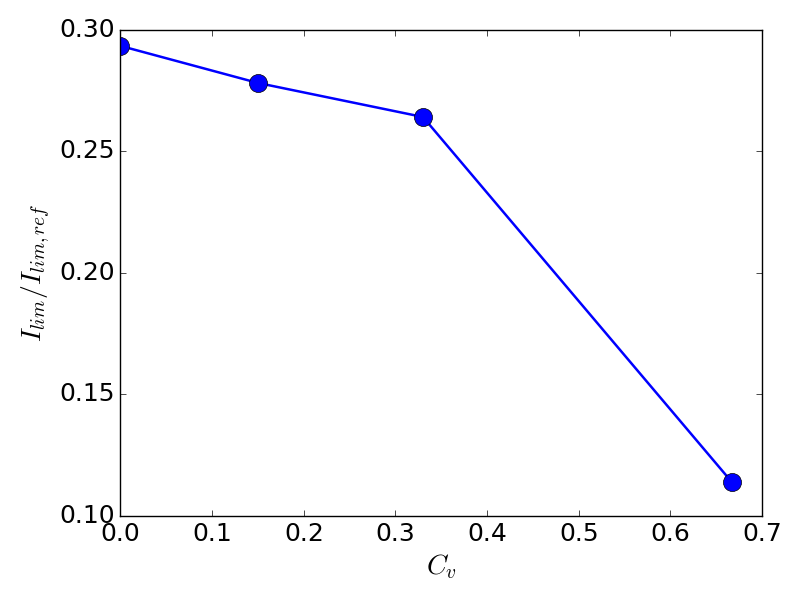}}
\caption{Normalized limiting current versus $C_v$ for nonhierarchical networks with $p=0.35$. The limiting currents are normalized by the limiting current associated with the network with $p=0$ and $C_v=0$, where all pores have the same diameter equal to 600 nm.}
\label{fig:ilim_smr_nonhierarch}
\end{figure}
Figure (\ref{fig:ilim_smr_nonhierarch}) depicts the variation of limiting current ($I_{\text{lim}}$) with increasing $C_v$. The limiting currents are normalized by a reference $I_{\text{lim}}$ computed for the network with $p=0$ and $C_v=0$. This plot reveals that the system conductance due to geometric modification is reduced by about a factor of 3 as $C_v$ increases.

\begin{figure}[H]
        \centering
        \begin{subfigure}[H]{0.5\textwidth}
                \includegraphics[width=\textwidth]{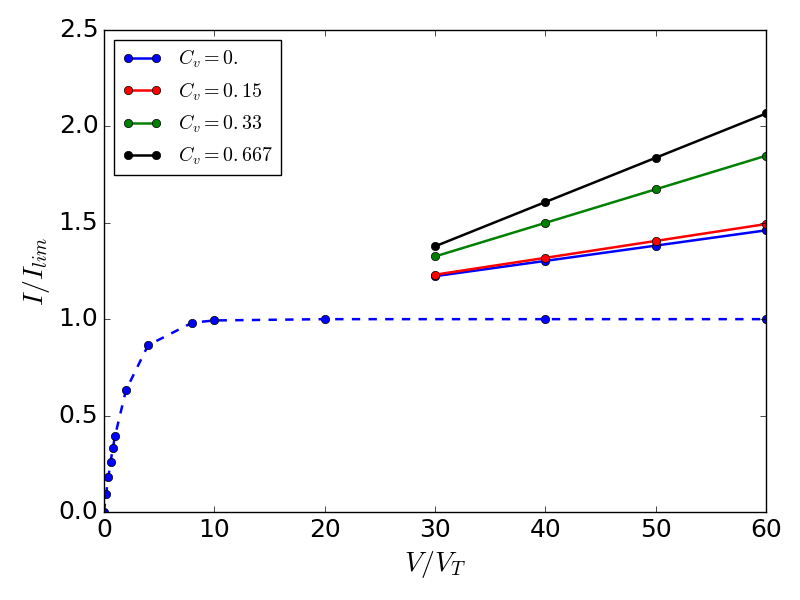}
                \caption{}
                \label{fig:iv_smr}
        \end{subfigure}%
        ~
        \begin{subfigure}[H]{0.5\textwidth}
                \includegraphics[width=\textwidth]{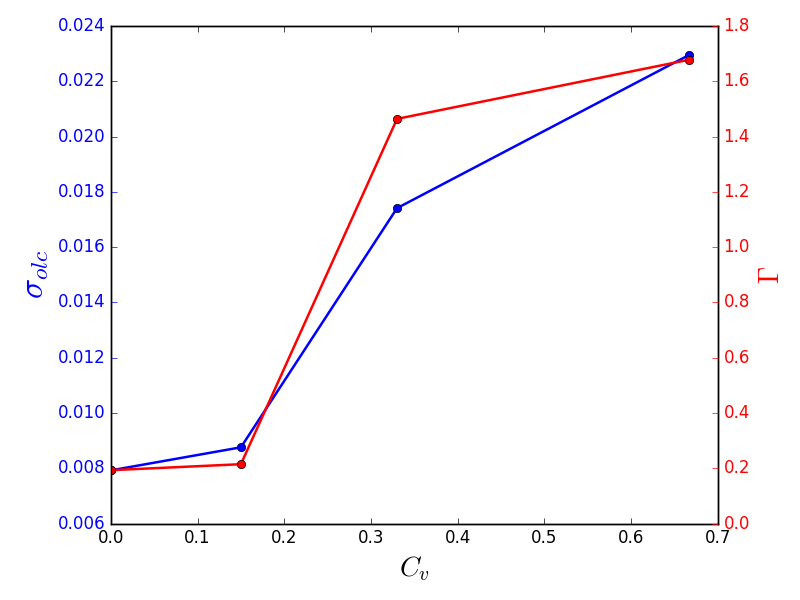}
                \caption{}
                \label{fig:sigma_smr}
        \end{subfigure}
        ~ 
         \caption{(a) The I-V curves of nonhierarchical random networks ($p=0.35$, $C_s\simeq0.01$, $\lambda_D\simeq0.008$, $\kappa=0.5$)  for different $C_v$. (b) The variation of overlimiting conductance (blue) and $\Gamma=\frac{\sigma_{\text{olc}, \text{adv}}}{\sigma_{\text{olc}, \text{sc}}}$ (red) vs. $C_v$.}
 \label{fig:nonhierarch_iv_smr}
 \end{figure}

Figure (\ref{fig:iv_smr}) shows the I-V curve for each network with different value of $C_v$. All I-V curves are normalized by their corresponding limiting currents. Larger values of $C_v$ result in higher overlimiting currents. We computed the overlimiting conductance for each I-V curve, as plotted in Figure (\ref{fig:sigma_smr}). Note that the conductance is computed in units of $I_{\text{lim}}/V_T$. As one can see in Figure (\ref{fig:sigma_smr}), there is an increasing trend in the overlimiting conductance of the system by increasing $C_v$. More interestingly, the red curve associated with $\Gamma$ in Figure (\ref{fig:sigma_smr}) indicates that increasing the pore size variation improves the contribution of advection to overlimiting  transport. This observation demonstrates that the pore size variation is a key factor for stimulating competitions between different modes of flow and the generation of internal flow loops. 

As $C_v$ reduces to zero, the network converges to a structure with uniform pore sizes. Our simulations indicate that in that limit, the mean flow inside each pore becomes extremely small (Pe<1e-5). However, Figure (\ref{fig:sigma_smr}) indicates that  even in that limit advection has finite contribution to overlimiting conductance. As explained in Section (2.5) and Figure (3a), the mechanism of advection in this case is due to internal recirculations within each pore while the net flow remains zero. Despite being 1D, our computational model captures these recirculations by incorporating the effects of non-uniformity of velocity and concentration profiles at each cross-section. 

\begin{figure}[H]
\makebox[\textwidth][c]{\includegraphics[width=1.1\textwidth]{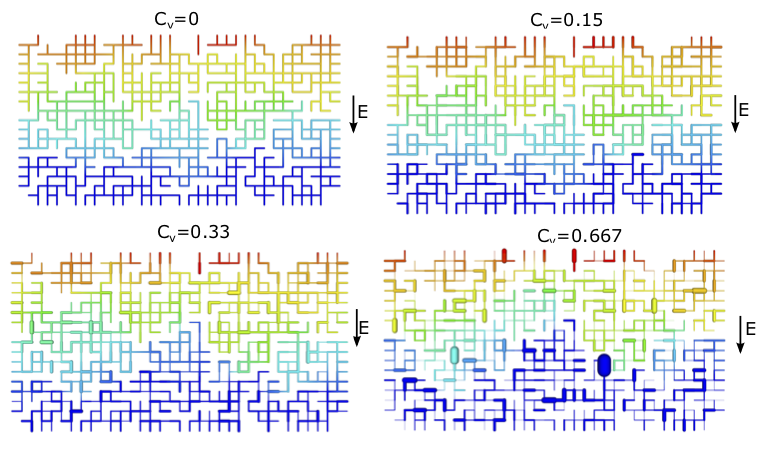}}
\caption{The steady state concentration fields in nonhierarchical networks of pores with different $C_v$. Increasing pore size variation stimulates generation of internal flow loops that move the high conductivity fluid in parallel in the direction of the applied field allowing higher charge transport.}
\label{fig:cbar_smars}
\end{figure}

Figure (\ref{fig:sigma_smr}) also indicates that the effects of internal recirculation within each pore on the overlimiting current is small and includes only $20\%$ of the effect of surface conduction. However, when pore sizes are random ($C_v>0$) and there are flow loops over the network of multiple pores, the advection effect can become significant compared to surface conduction. 
The steady state concentration fields corresponding to different values of $C_v$ are depicted in Figure (\ref{fig:cbar_smars}). The results are obtained for $V/V_T=40$. By increasing $C_v$, the porous network involves wider range of pore dimeters which allows the generation of flow loops. As demonstrated in Figure (\ref{fig:sigma_smr}), these flow loops can contribute to the transport of charge by redistributing the ICP zone, such that ``more parallel'' high conductivity zones are available for charge transport. For the network with uniform structure ($C_v$=0) the shock front looks uniform in the horizontal direction and driven primarily by surface conduction.

\subsection{Pore blocking probability}
We explored the impact of pore blocking probability on the I-V curve of nonhierarchical networks. Similar to previous section, we considered networks with mean pore diameter of 600 nm, uniform surface charge density $\sigma=-10 mC/m^2$, and reservoir concentration of $C_{\text{ref}}=50 mM$, fixed $C_v=0.33$. The steady state behavior of the system is computed considering a range of pore blocking probability, $p=\{0., 0.1, 0.2, 0.3, 0.4, 0.5\}$. The $C_s$ values for different pore blocking probabilities are close and equal to $C_s\simeq 0.012$, which corresponds to $h_p\simeq 167$ nm and $\lambda_D=0.008$. We here report the results ensemble averaged over three simulations conducted for each value of $p$. We observed about 2\% variation in $I/I_{\text{lim}}$ with respect to the ensemble averaged I-V curves. 

\begin{figure}[H]
\makebox[\textwidth][c]{\includegraphics[width=0.55\textwidth]{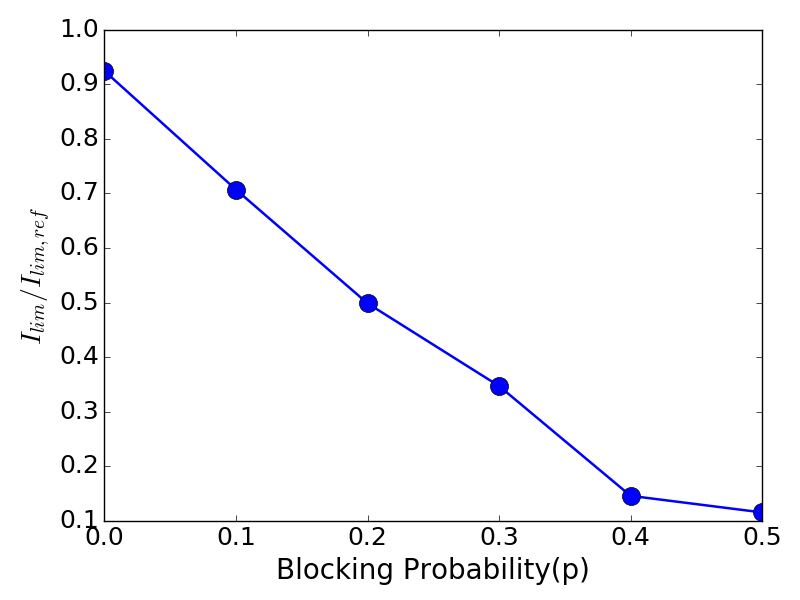}}
\caption{Normalized limiting current versus pore blocking probability for nonhierarchical networks with $C_v=0.33$. The currents are normalized by the limiting current of the network constructed by $p=0$ and $C_v=0$, where all pores have the same diameter equal to 600 nm.}
\label{fig:ilim_p_nonhierarch}
\end{figure}

For each value of $p$, the I-V curve is computed in the overlimiting regime and normalized by corresponding $I_{\text{lim}}$. Figure (\ref{fig:ilim_p_nonhierarch}) depicts the variation of $I_{\text{lim}}$ with respect to $p$. Blocking more pores has led to a significant reduction in the limiting current, which corresponds to the reduction in the geometric conductivity of the network. This implies that blocking more pores reduces the number of pathways for transport of charge towards the membrane. 

\begin{figure}[H]
        \centering
        \begin{subfigure}[H]{0.5\textwidth}
                \includegraphics[width=\textwidth]{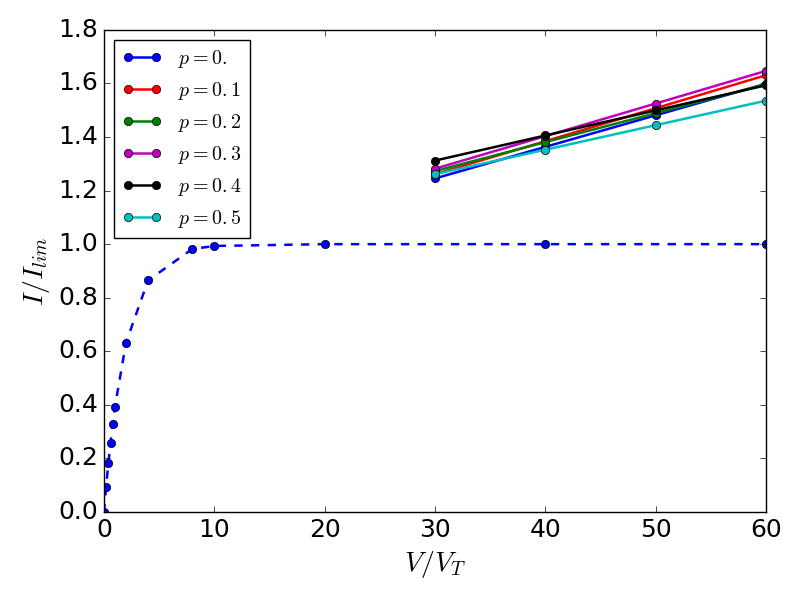}
                \caption{}
                \label{fig:iv_p}
        \end{subfigure}%
        ~
        \begin{subfigure}[H]{0.5\textwidth}
                \includegraphics[width=\textwidth]{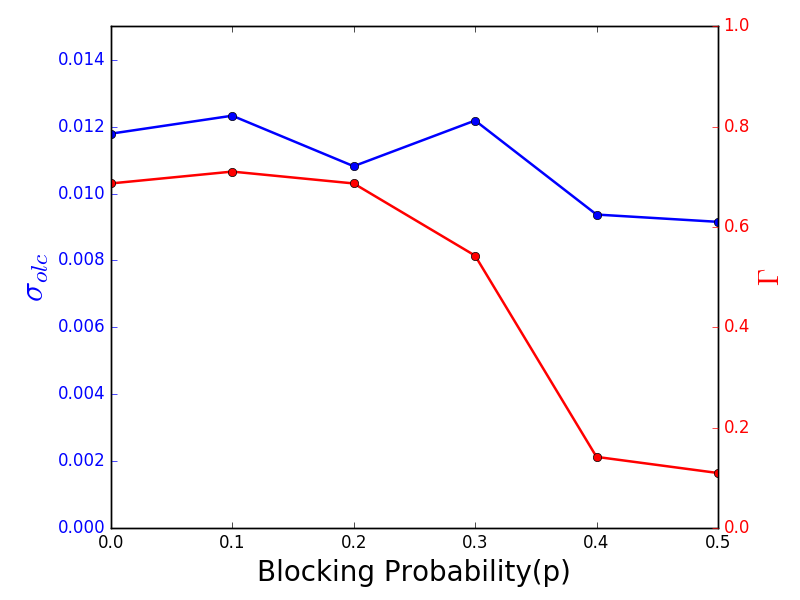}
                \caption{}
                \label{fig:sigma_p}
        \end{subfigure}
        ~ 
         \caption{(a) The I-V curves of nonhierarchical networks for different values of blocking probability and $C_v=0.33$, $\lambda_D\sim0.008$, $C_s\sim0.012$, and $\kappa=0.5$. (b) The overlimiting conductance (blue curve) and $\Gamma$ (red curve) vs. pore blocking probability.}
 \label{fig:nonhierarch_iv_p}
 \end{figure}

\begin{figure}[H]
\makebox[\textwidth][c]{\includegraphics[width=1.15\textwidth]{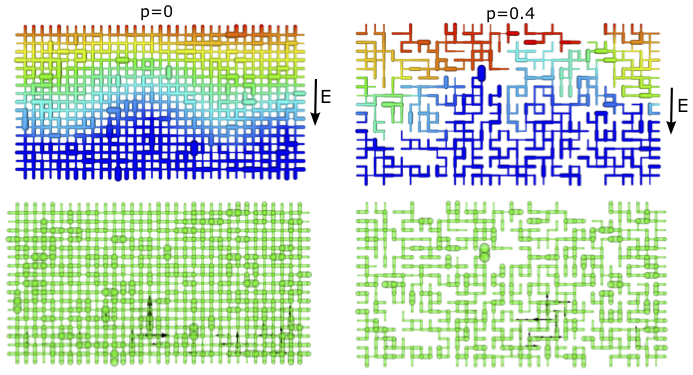}}
\caption{The steady state concentration and the flow fields in the nonhierarchical networks for two values of pore blocking probability, p=0 (left) and p=0.4 (right). Large values of p reduces the number of loops in the network geometry, leading to the exclusion of induced flow loops in the network.}
\label{fig:cbar_flow_p0_04}
\end{figure}

The I-V characteristics of the considered networks are shown in Figure (\ref{fig:iv_p}). The impact of $p$ on the overlimiting response seems to be small. Figure (\ref{fig:sigma_p}) reveals that the overlimiting conductance of the system is not sensitive to the variation of $p$. The dominant mechanism in this geometry is surface conduction, and thus the variation of blocking probability does not significantly affect the overlimiting current of the system. On the same plot, the variation of $\Gamma$ vs. $p$ confirms that the contribution of advection mechanism to charge transport is smaller than the surface conduction, and the contribution decreases as the pore blocking probability increases. We attribute this decrease to be due to the reduction of possible flow loops when more pores are blocked. 
Figure (\ref{fig:cbar_flow_p0_04}) depicts the steady state concentration fields and the flow fields of the nonhierarchical networks corresponding to $p=0$ and $p=0.4$ at $V/V_T=50$. In all networks there are induced flow loops that are located close to the cation-selective membrane. The induced flow is much smaller further away from the membrane. The length scale of the flow loops are usually at the order of pore's length or comparable to the length scale of the depletion zone. Blocking more pores reduces the presence of physical loops in the network geometry, leading to the generation of fewer flow loops. These flow loops are active in small zones inside the depleted region.

\section{Hierarchical Networks: Dependency on Network Parameters} \label{sec:hierarch_pore_size_p}
\subsection{Pore size variability}
Similar to what we did for nonhierarchical networks, we computed the steady state response of hierarchical networks for different values of $C_v$. For all simulations we assume uniform surface charge density $\sigma=-10 mC/m^2$, reservoir concentration of $C_{\text{ref}}=50 mM$, and pore blocking probability $p=0.35$. To be able to assess the impact of pore size randomness on the advection mechanism, we ran our simulations in the small surface conduction regime. We compare our numerical results for hierarchical networks to those computed for nonhierarchical networks with the same $C_v$. To reduce the statistical variations, we repeated our simulations for each set of network parameters three times and report the ensemble averaged results. We observed statistical variations of about 6\% in $I/I_{\text{lim}}$ relative to the ensemble averaged I-V curves. 
We considered three values of $C_v=\{0, 0.33, 0.667\}$ for this investigation. The effective $h_p$ of the networks considered are $h_p=\{ 150, 170, 220\}$ nm, which corresponds to $\lambda_D=\{ 0.009, 0.008, 0.006\}$ and $C_s=\{0.013, 0.012, 0.009\}$. 
\begin{figure}[H]
        \centering
        \begin{subfigure}[H]{0.4\textwidth}
                \includegraphics[width=\textwidth]{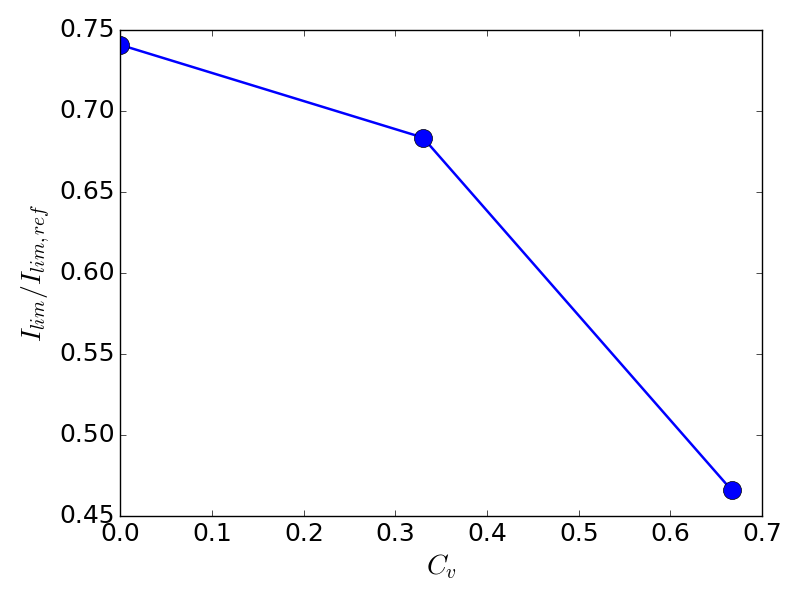}
                \caption{}
                \label{fig:ilim_smr_hierarch}
        \end{subfigure}%
        ~
        \begin{subfigure}[H]{0.4\textwidth}
                \includegraphics[width=\textwidth]{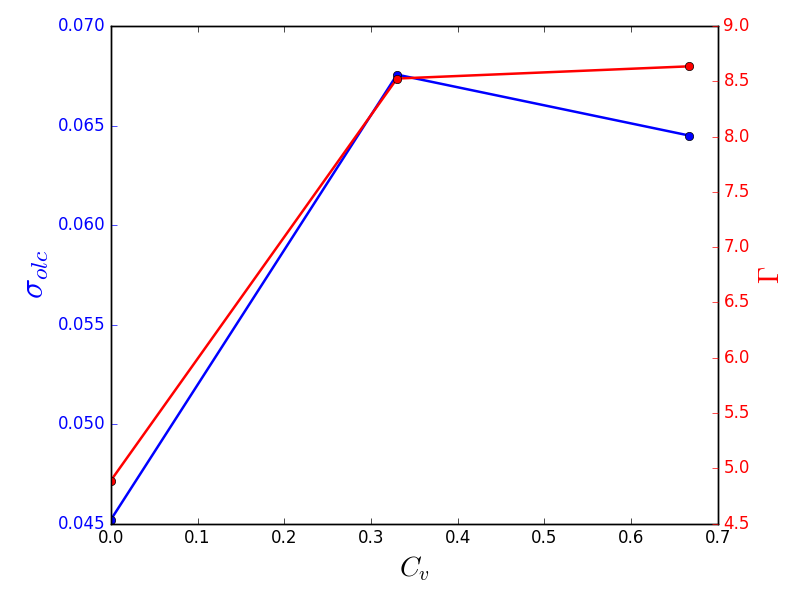}
                \caption{}
                \label{fig:olc_sigma_smr_hierarch}
        \end{subfigure}
          \begin{subfigure}[H]{0.55\textwidth}
                \includegraphics[width=\textwidth]{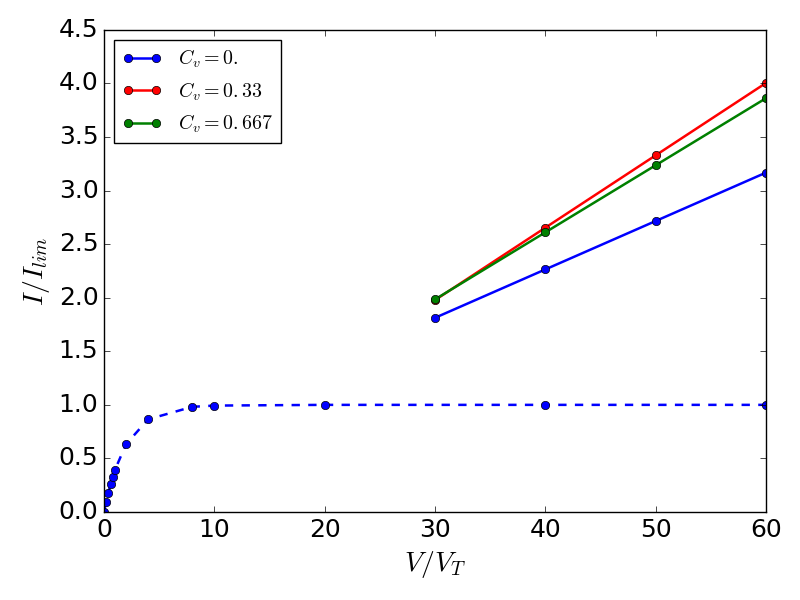}
                \caption{}
                \label{fig:iv_smr_hierarch}
        \end{subfigure}
         \caption{(a) Normalized limiting current vs. pore size variation ($C_v$) for hierarchical random networks with $\kappa=0.5$, $p=0.2$ and over all $C_v$ values $\lambda_D\sim0.008$ and $C_s\sim0.012$ are maintained fairly constant. (b) The overlimiting conductance (blue curve) and $\Gamma$ (red curve) vs. $C_v$. All $\sigma_{\text{olc}}$ are in units of $I_{\text{lim}}/V_T$. (c) Normalized I-V curves used for the calculation of $\sigma_{olc}$.}
 \label{fig:hierarch_olc_sigma_iv_smr}
 \end{figure}
Figure (\ref{fig:ilim_smr_hierarch}) depicts the variation of limiting current with $C_v$ for hierarchical networks. Figure (\ref{fig:iv_smr_hierarch}) presents the I-V characteristics normalized by $I_{\text{lim}}$. Similar to the trends observed earlier, increasing pore size variation increases the normalized current through the system. The overlimiting conductance and the values of $\Gamma$ are plotted  in Figure (\ref{fig:olc_sigma_smr_hierarch}). The results indicate that the contribution of advection mechanism is more significant in the hierarchical configuration than the nonhierarchical setting leading to higher charge transport through the network. As one example, we present the concentration and flow field of the hierarchical network constructed for $C_v=0.667$ in Figure (\ref{fig:smr_hierarch}) when $V/V_T=40$. This network has depicted the highest contribution of advection to OLC. The backward pressure-driven flow has led to the propagation of the depletion zone further away from the membrane. The induced flow loops on the left side of the network facilitate the transport of higher charge towards the membrane, leading to a higher I-V response.\\
\begin{figure}[H]
        \centering
        \begin{subfigure}[H]{0.54\textwidth}
                \includegraphics[width=\textwidth]{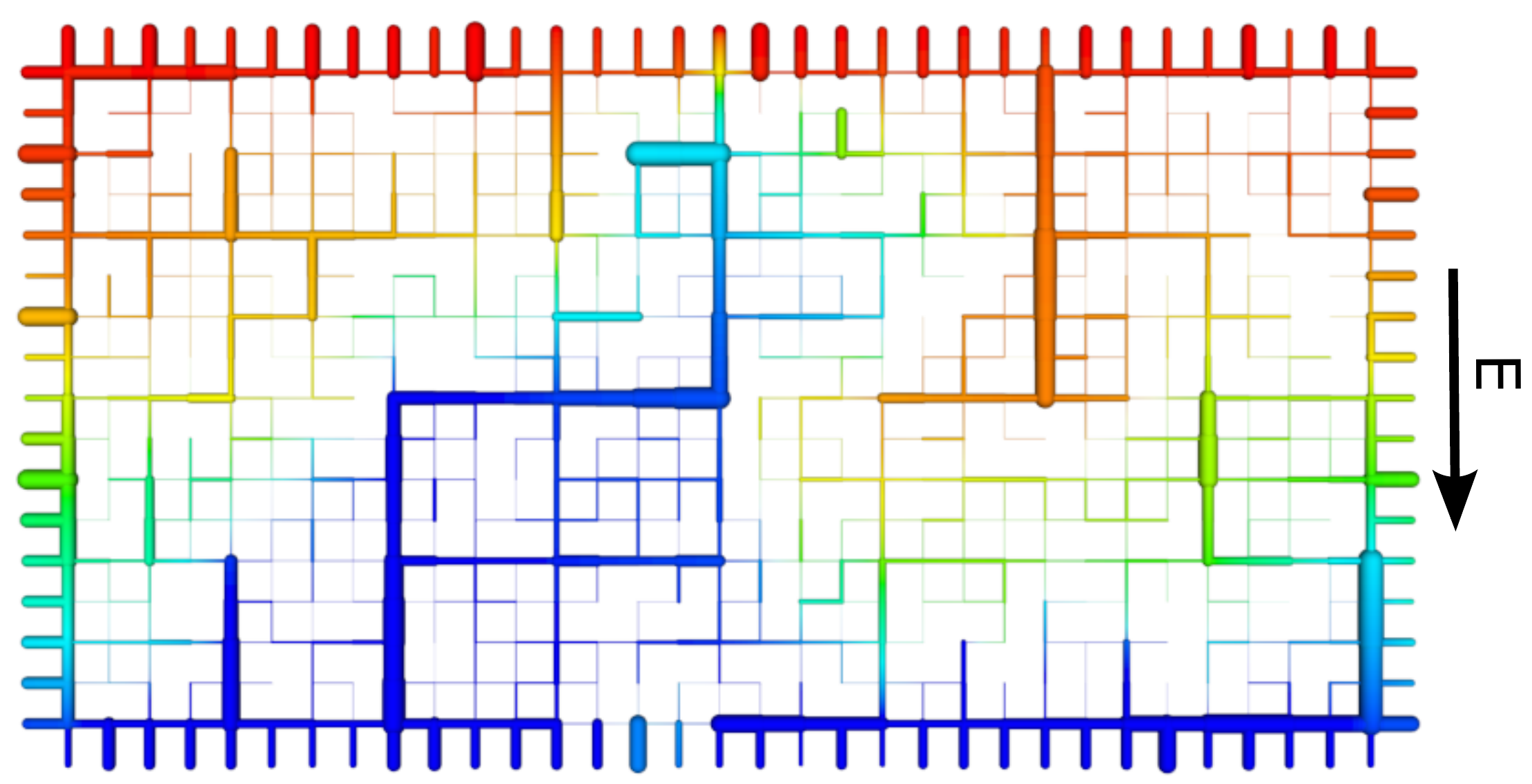}
                \caption{}
                \label{fig:cfield}
        \end{subfigure}%
        ~ 
        \begin{subfigure}[H]{0.5\textwidth}
                \includegraphics[width=\textwidth]{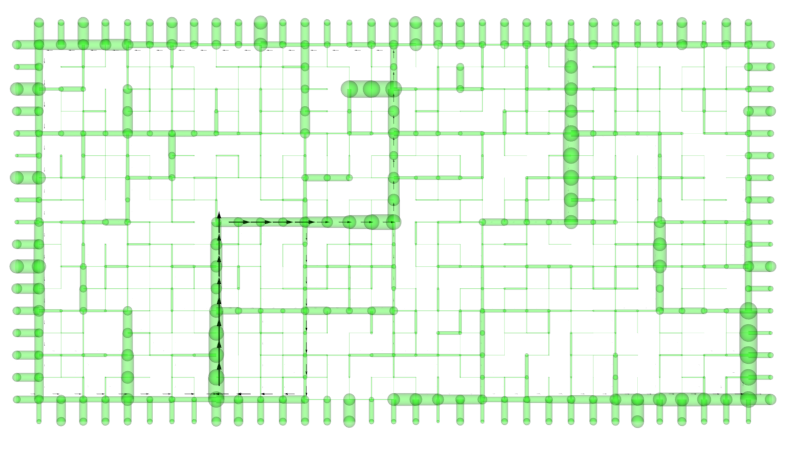}
                \caption{}
                \label{fig:flowfield}
        \end{subfigure}
        ~ 
         \caption{(a) The steady state concentration field in the hierarchical network with $C_v$ = 0.667. (b) The induced flow field in the network.}
 \label{fig:smr_hierarch}
 \end{figure}

\subsection{Pore blocking probability}
We computed the steady state responses of hierarchical networks with various pore blocking probability $p=\{0, 0.2, 0.4\}$. For each value of $p$, we design a hierarchal network that has the same surface charge density and effective $h_p$ as the nonhierarchical network generated with the same value of $p$. This ensures that the two networks have the same surface conduction strength, $C_s\simeq0.012$. Additionally, parameters $\lambda_D\simeq0.008$, and $C_v=0.33$, match those in the nonhierarchical counterparts. This allows comparisons determining the impact of network hierarchy on the overlimiting transport. For dimensional parameters we considered $C_{\text{ref}}=50 mM$, and mean pore diameter of the coarsest scale equal to $2 \mu m$. The effective pore size of the networks are close, $h_p\simeq169$ which is close to the one associated with the nonhierarchical networks. The results presented here are ensemble averaged over three simulations for each values of $p$ considered in this study. The statistical variations in $I/I_{\text{lim}}$ observed for different values of $p$ are in the range of 8\% to 15\% of the ensemble averaged I-V curves.
\begin{figure}[H]
        \centering
        \begin{subfigure}[H]{0.5\textwidth}
                \includegraphics[width=\textwidth]{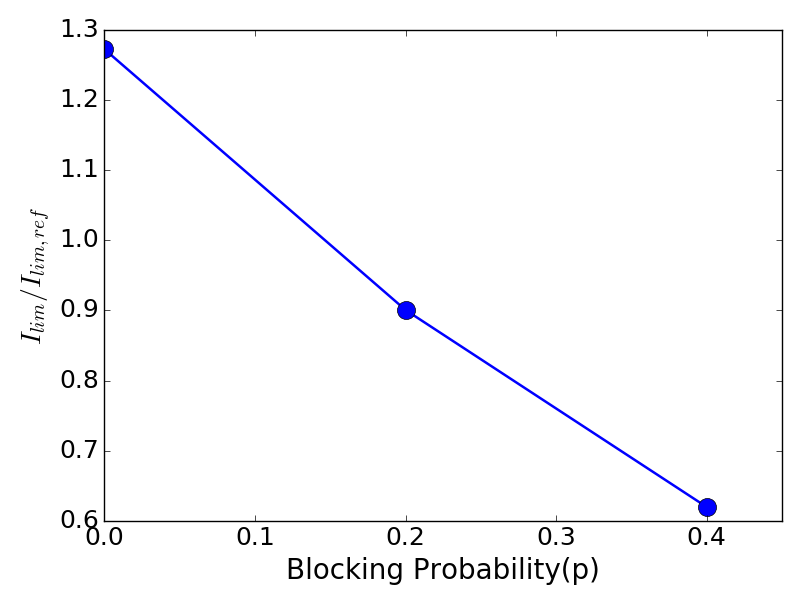}
                \caption{}
                \label{fig:ilim_p_hierarch}
        \end{subfigure}%
        ~
        \begin{subfigure}[H]{0.5\textwidth}
                \includegraphics[width=\textwidth]{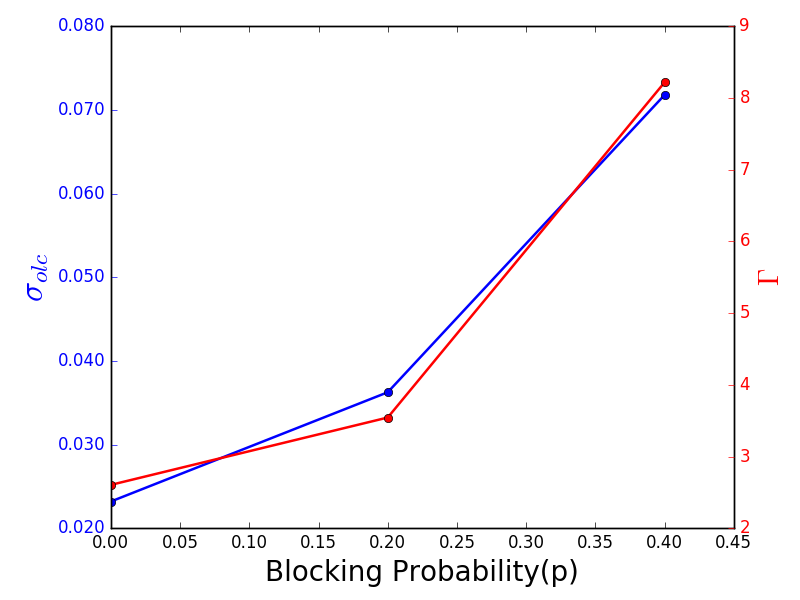}
                \caption{}
                \label{fig:olc_sigma_p_hierarch}
        \end{subfigure}
          
          \begin{subfigure}[H]{0.6\textwidth}
                \includegraphics[width=\textwidth]{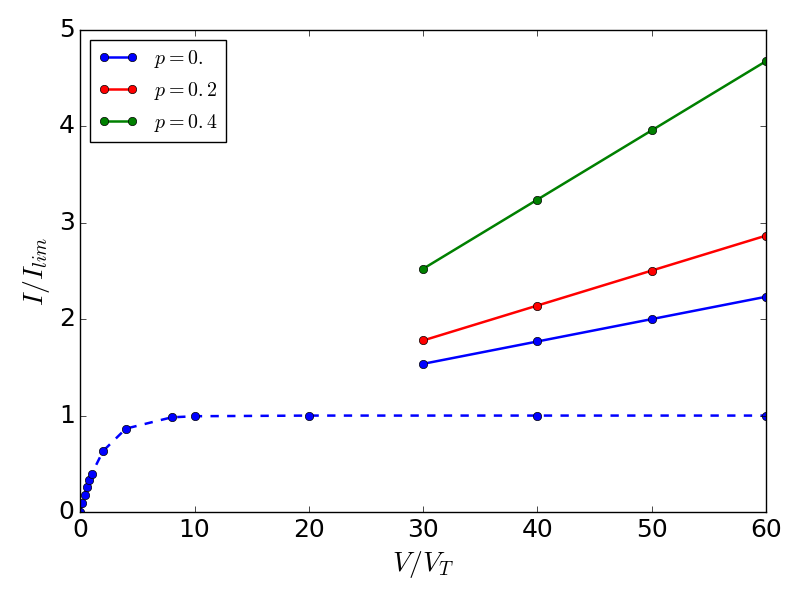}
                \caption{}
                \label{fig:iv_p_hierarch}
        \end{subfigure}
         \caption{(a) Normalized limiting current vs. pore blocking probability ($p$) for hierarchical networks with $C_v=0.33$, $C_s\sim 0.012$, $\lambda_D\sim 0.008$ and $\kappa=0.5$. (b) The variation of overlimiting conductance and $\Gamma$ with respect to $p$ (c) the I-V curves of hierarchical networks for different values of $p$.}
 \label{fig:hierarch_olc_sigma_iv_p}
 \end{figure}
 As shown in Figure (\ref{fig:ilim_p_hierarch}), similar to nonhierarchical networks, the limiting current of the hierarchical networks decreases by increasing the pore blocking probability. The overlimiting conductance and the corresponding $\Gamma$ are shown in Figure (\ref{fig:olc_sigma_p_hierarch}). Figure (\ref{fig:iv_p_hierarch}) depicts the normalized I-V curves of hierarchical random networks. As opposed to nonhierarchical geometry, there is a noticeable increase in the normalized overlimiting conductance by increasing $p$ up to 0.4. 
\begin{figure}[H]
        \centering
        \begin{subfigure}[H]{1\textwidth}
                \includegraphics[width=\textwidth]{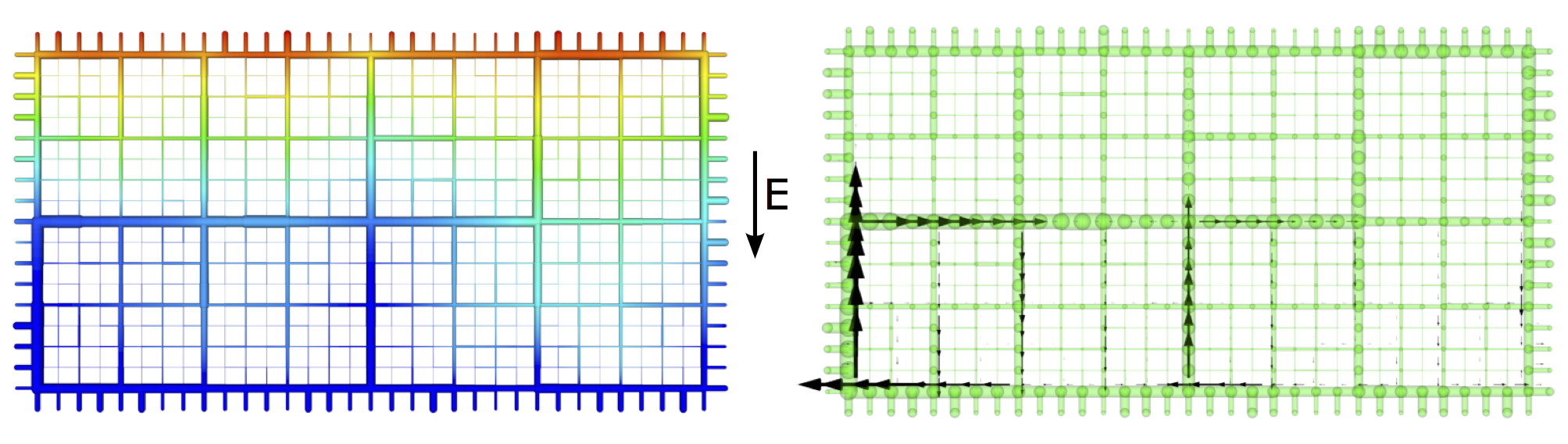}
                \caption{}
        \end{subfigure}%
          
        \begin{subfigure}[H]{1\textwidth}
                \includegraphics[width=\textwidth]{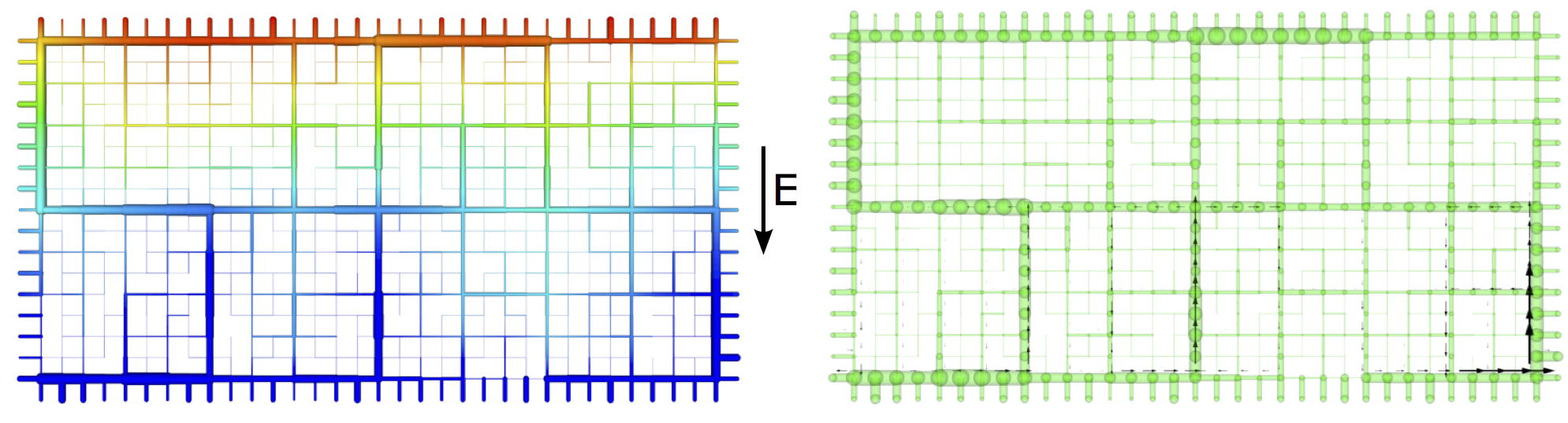}
                \caption{}
        \end{subfigure}
          
          \begin{subfigure}[H]{1\textwidth}
                \includegraphics[width=\textwidth]{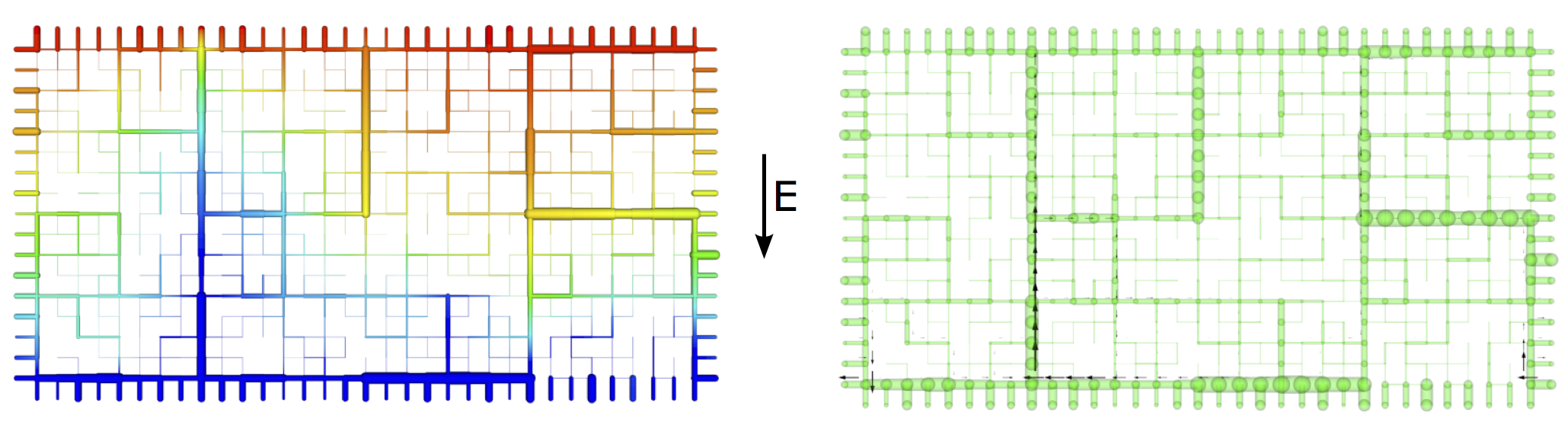}
                \caption{}
        \end{subfigure}

         \caption{The steady state concentration fields and the corresponding flow fields at $V/V_T=40$ for (a) p=0, (b) p=0.2, and (c) p=0.4. All networks involve flow loops that contribute to salt transport, and results in $\Gamma>1$. The network topology in (c) has induced a large flow loop at the scale of the domain in the vertical direction. This loop allows transport of higher amount of salt from the top region towards the membrane, while assisting the propagation of deionization shock towards the reservoir.}
 \label{fig:p_hierarch}
 \end{figure}
Considering the results of Figure (\ref{fig:hierarch_olc_sigma_iv_smr}) and (\ref{fig:hierarch_olc_sigma_iv_p}), we can conclude that over wide range of geometries considered, the hierarchical structures sustain much larger advective transport compared to the nonhierarchical ones. In the range of $C_s\sim 0.01$ a hierarchical structure can sustain a very large overliminiting current dominated by advection while a nonhierarchical structure is mostly governed by the surface conduction mechanism and much lower transport. 

The steady state concentration and the flow fields of hierarchical networks corresponding to p=$\{0, 0.2, 0.4\}$ are shown in Figure (\ref{fig:p_hierarch}). These are numerical results for $\Delta V/V_T = 40$. In all three geometries, the induced flow loops have formed regions of desalinated water with the same length scales as the length scales of the loops. The network with $p=0.4$ reveals the largest expansion of the deionized region, which has been occurred on the left side of the system, where there is a strong backward pressure-driven flow pushing the shock front towards the reservoir.